\documentclass[journal,onecolumn,draftclsnofoot]{IEEEtran} 

\usepackage{calc}
\newcommand{\figwidth}{\minof{\columnwidth}{5.0in}}

\usepackage{cite}
\usepackage{graphicx}
\graphicspath{{./../fig/}}

\ifCLASSINFOpdf
  \usepackage{graphicx}
  \graphicspath{{./../fig/}}
\else
\fi

\usepackage[cmex10]{amsmath}
\usepackage{multirow}
\usepackage{booktabs}

\begin{document}
\title{Capacity Considerations for Secondary Networks in TV White Space}

\author{Farzad~Hessar,~\IEEEmembership{Student Member,~IEEE,}
				Sumit~Roy,~\IEEEmembership{Fellow,~IEEE}
\\ EE. Department, University of Washington
\\ \{farzad, sroy\}@u.washington.edu
}

\maketitle
\thispagestyle{empty}

\begin{abstract}
The so-called `TV white spaces' (TVWS) - representing unused TV channels in any given location as the result of the transition to digital broadcasting - designated by U.S. Federal Communications Commission (FCC) for unlicensed use \cite{Std:FCC,Std:FCC10,Std:FCC12} presents significant new opportunities within the context of emerging 4G networks for developing new wireless access technologies that meet the goals of the US National Broadband Plan\cite{Rp:NBP} (notably true broadband access for an increasing fraction of the population). There are multiple challenges in realizing this goal; the most fundamental being the fact that the available WS capacity is currently not accurately known, since it depends on a multiplicity of factors - including system parameters of existing incumbents (broadcasters), propagation characteristics of local terrain as well as FCC rules.
In this paper, we explore the capacity of white space networks by developing a detailed model that includes all the major variables, and is cognizant of FCC regulations that provide constraints on incumbent protection.  Real terrain information and propagation models for the primary broadcaster and adjacent channel interference from TV transmitters are included to estimate their impact on achievable WS capacity. The model is later used to explore various trade-offs between network capacity and system parameters and suggest possible amendments to FCC's incumbent protection rules in the favor of furthering white space capacity.
\end{abstract}


\IEEEpeerreviewmaketitle

\section{Introduction}
Next generation of cellular data networks will face an exponential growth of wireless data traffic resulting from the boom in multimedia applications running on smart phones, tablets, and other wireless devices \cite{ele:Cisco}. Available 4G (licensed) spectrum will clearly be insufficient to meet this demand, leading to cellular operators searching for novel mechanisms to achieving operational efficiencies that enhance network capacity. Obviously, an essential response to this increasing demand is the opening of new spectrum - both licensed, and recently, {\em unlicensed}. Resulting from the transition to digital TV broadcasting \footnote{In the U.S. this was completed by June 2009.} and the consequent freeing up of VHF/UHF spectrum  (between 50-700 MHz),  FCC took the unprecedented step of allocating significant portions for unlicensed use, intended for providing enhanced wireless broadband access \cite{Std:FCC10, Std:FCC12}. These bands - collectively denoted as TV White Spaces (TVWS) - are interspersed with the 4G 700 MHz {\em licensed} bands \footnote{For example, portions of 698-806 MHz  were auctioned off by 2008 in the U.S. to provision for 4G mobile broadband services.} and will allow secondary (unlicensed) users to opportunistically access them provided {\em interference protection guarantees} to the neighboring primary (licensed) networks are ensured. A strategy for {\em coordinated use of both licensed and unlicensed 700 MHz spectrum} is the likely answer for 4G cellular operators, just as offloading to 802.11 WLAN hotspot networks has proved to be a boon for 3G network providers. TVWS (sometimes subbed as `super Wi-Fi') may potentially provide even more significant offload/spectrum aggregation opportunities, considering other proximal government held spectrum are also being explored for de-regulation \cite{PCAST, NTIA1, NTIA2, NTIA3, TRID}.

Clearly, the most critical task for a Dynamic Spectrum Access (DSA)-based cognitive users is finding spectrum `holes' efficiently \cite{paper:roy12}, i.e., spectrum resources in {\em time-frequency-spatial} dimensions at any location that are currently available, which can be then used for unlicensed operation. Current FCC rules of operation for unlicensed users in TVWS require them to register with and obtain recommendations from a list of approved Database Administrators (DBA) - such as a list of available TVWS channels - for their operation, so as to ensure interference protection to the incumbents. The DBA is responsible for complying with FCC regulations \cite{Std:FCC12} in modeling all primary users status, as a basis of providing the necessary recommendations to the secondary users requesting access.

In this work, we explore a fundamental issue pertaining to WS usage by secondary networks, that may be succinctly captured by the
question {\em How much WS network capacity actually exists per FCC rules}?. Clearly, the answer to this question is of paramount importance to 4G network operators as they consider new infrastructure for unlicensed WS access as part of their operations. We show that that it depends on multiple factors:
\renewcommand{\labelitemi}{$\diamond$}
\begin{itemize}
	\item Primary Network Parameters (transmit power and signal masks, modulation/coding)
	\item FCC rules for protection of primary by limiting secondary operation (protection regions, adjacent TV channels, primary receiver design and sensitivity)
	\item Propagation Characteristics (location dependent terrain models, heights etc.)
	\item Secondary network parameters (transmit power and signal masks, modulation/coding, multiple access schemes)
\end{itemize}
Prior work on this topic, notably \cite{paper:Markis, conf:sahai2} does not provide a sufficiently nuanced exploration of this question (available secondary network capacity) as a function of {\em all} the parameters above. Notably, the {\em structure of the spatial variations} in secondary capacity has not been adequately captured, in our opinion. Further, FCC regulations have evolved significantly since the first release of TVWS \cite{Std:FCC,Std:FCC12} which in turn impact TVWS capacity analysis. Mutual effects of secondary and primary networks, such as co-channel and adjacent-channel interference, are not considered. Furthermore, realistic and empirical path loss models that are based on actual terrain information together with sensitivity to path loss variation have not been previously used.

We develop a {\em spatial description of WS capacity} via a model that captures both primary and secondary network aspects as well as channel and environmental characteristics. An important side benefit of our analysis is the spotlight it shines on {\em whether the  current incumbent protection rules proposed by the FCC may need amending in the interests of promoting more WS availability}. Our analysis thus provides fundamental insights into aspects of coexistence between secondary users and primary transmitters as a function of FCC regulations \cite{Std:FCC10,Std:FCC12}. The rest of the paper is organized as follows. Section II defines the network structure for coexistence of secondary and primary users. Section III formulates capacity of each secondary cell and summarizes the relevant current FCC rules. Section IV discusses propagation (path loss) models and the underlying physical environmental parameters while the interference model are introduced in Sec. V. Numerical calculations  in Sec. VI provides high level results on available secondary capacity whereas Section VII explores multiple trade-offs that arise. Sec. VIII concludes the paper and some supplementary details are related to Appendices A and B.

\section{Secondary Network Architecture}
\label{sec:CellularNetworkStructure}

\begin{figure}[!t]%
	\centering
	\includegraphics[width=\figwidth]{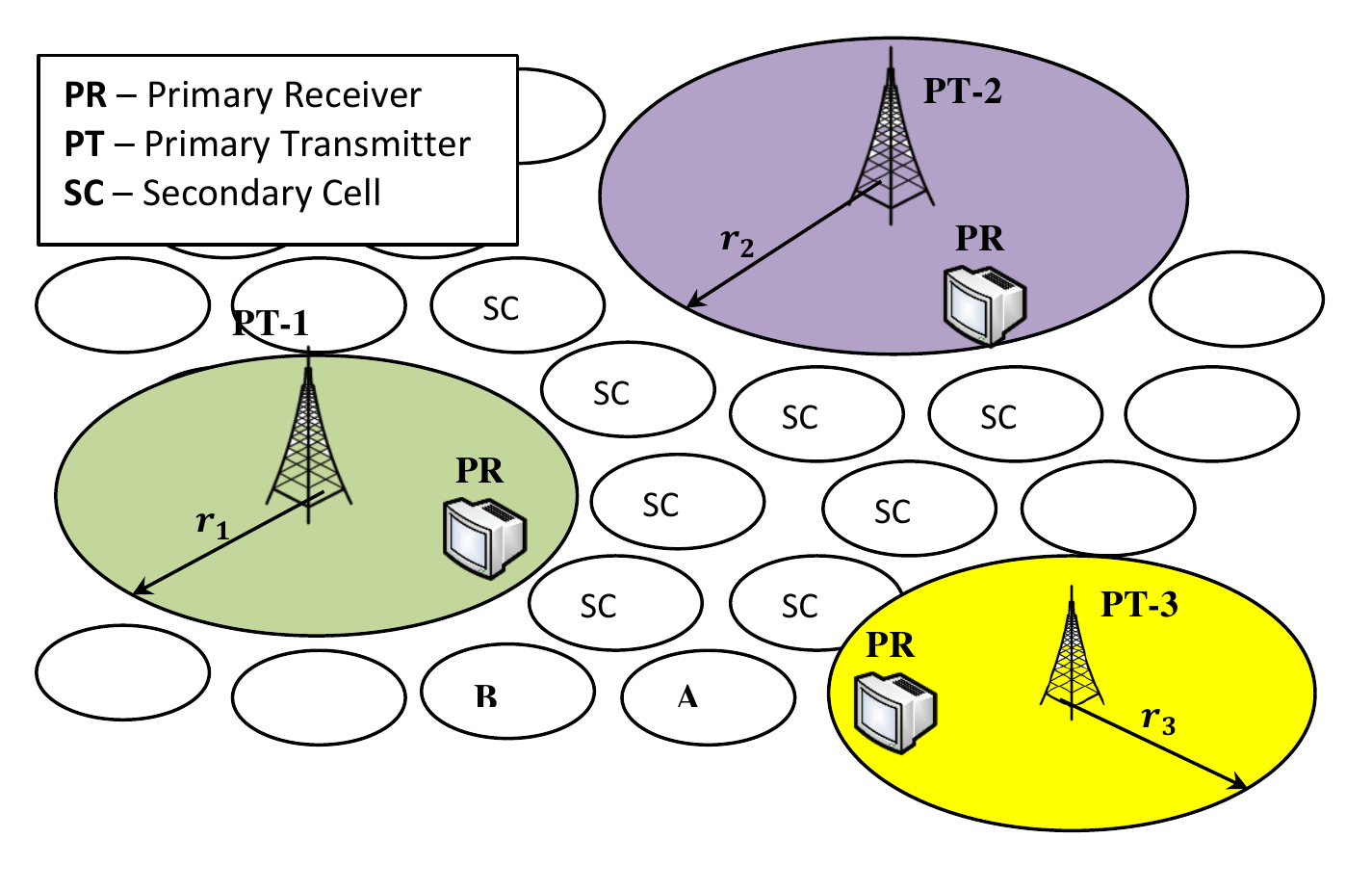}
	\caption{Coexistence of primary and secondary cellular networks}%
	\label{fig:cellStruct}%
\end{figure}

Cellular communication systems are based on the notion of {\em frequency reuse} which allows a channel to be spatially re-used by different users, as long as the co-channel interference is within acceptable bounds. However, for TVWS applications, the cellular layout of the secondary cells (SC) is further restricted by the primary protection regions, as shown in Fig. \ref{fig:cellStruct}. This figure presents TV towers as an {\em irregular} primary network, where each primary cell corresponds to the coverage area of the associated tower. Here, $r_i$ is the maximum distance at which the received TV signal is received above the detection threshold \cite{FCC:LongleyRice, Std:FCC12,recom:itu}. The regions outside the primary transmitter coverage area constitute the white/gray space \cite{Rp:Peha11} that can be utilized by secondary networks; secondary cells are naturally much smaller than primary cells, due to power and antenna height limitations.

Characterizing the white spaces and the primary-to-secondary interference requires knowledge of deployment of TV towers (primary transmitters) and their parameters, which is available from the FCC database \cite{FCC:techrep1}. Comparing the examples of  TV channels 13 (see Fig. \ref{fig:Ch13Pr}) and 5 (see Fig. \ref{fig:Ch5Pr}) indicates that the patterns of spatial reuse for primary (TV channels) are irregular (across channels). As a result, the availability of WS shows significant spatial variations, which we explore in more details later.

Define the set of available channels in a secondary cell $A$ (according to FCC rules for protection of {\em primary users}) as
\begin{align}
\Upsilon(A, \Gamma_A) = \left\{c_i :\, \mbox{ Channel } c_i \mbox{ is available at cell A with parameter }\Gamma_A\right\} \nonumber
\end{align}
where $\Gamma_A$ represents all relevant network parameters. Note that $\Upsilon(B, \Gamma_B)$ represents a different list of available channels. In order to protect secondary users from secondary interference, only a subset of available channels are assigned to each secondary cell which is considered in section \ref{sec:InterferenceModel}.

\section{Secondary Link Capacity}
We begin developments by computing the Shannon capacity of a (hypothetical) secondary transmitter-receiver pair located at a point, i.e. the capacity of an infinitesimal cell with one active link where the separation between the source and receiver is negligible. Assuming that the secondary users are allocated {\em exactly one} of the available channels $c_i\in\Upsilon(A, \Gamma_A)$ at any secondary cell $A$, the capacity is a function of the usual parameters - {\em signal to noise + interference ratio} and {\em available bandwidth}
\begin{align}
C_{cell} =  W_0\,\log_2(1+\mbox{SINR})
\end{align}
where $W_0=6$ MHz represents the bandwidth of an NTSC TV channel. Since the allocated channel $c_i$ is not available at every location, a more appropriate measure of available capacity is an {\em area average}. To do this, we introduce a Bernoulli random process for the availability of any secondary WS channel, as follows:
\begin{align}
\label{eq:BWParams}
W(Q_T, \Gamma) = \left\{\begin{array}{cc}
		W_0 & \mbox{    } Q_T\not\in \Omega \mbox{ and } \Gamma \vdash \mbox{FCC rules}\\
		0   & \mbox{    } Q_T\in \Omega  \mbox{ \ or \ } \Gamma \not\vdash \mbox{FCC rules}
		\end{array}
		\right.
\end{align}
where $Q_T$ is the transmitter location, $\Omega$ is the set of all protection regions for the channel currently under exploration. $\Gamma$, as defined before, is the set of network parameters that can affect channel availability, including transmission power, antenna height above average terrain, out of band emission, etc., which must comply with FCC requirements ($\vdash$) for incumbent protection. Using this, define the {\em channel availability} $p(\Gamma) = Pr[Q_T\not\in \Omega]$ as the normalized average of $W(Q_T, \Gamma)$ over an area $\mathcal{A}$
\begin{align}
p(\Gamma)= \frac{1}{\mathcal{A}}\int_{\mathcal{A}}{\frac{W(Q_T, \Gamma)}{W_0}dQ_T}
\end{align}
The capacity averaged over transmitter and receiver locations is thus
\begin{align}
\label{eq:AvgChan}
\overline{C_{cell}}(\Gamma) = p(\Gamma)W_0\int_{\mathcal{A}}{\frac{\log_2(1+\mbox{SINR}(Q_R))}{\mathcal{A}}dQ_R}
\end{align}
where the SINR depends on (secondary) receiver's location $Q_R$. Eq. (\ref{eq:AvgChan}) can be calculated for every channel separately and the overall capacity of a cell that is exploiting all available channels is obtained by summing over all channels
\begin{align}
\label{eq:AvgChanSum}
\overline{C_{cell,total}}(\Gamma) = \sum_{c_i\in\Upsilon(cell, \Gamma_{cell})} {p(\Gamma,c_i)W_0\int_{\mathcal{A}}{\frac{\log_2(1+\mbox{SINR}(Q_R,c_i))}{\mathcal{A}}dQ_R}}
\end{align}

\subsection{Channel Availability}
Availability of every permissible TV channel $c_i\in\{2:51\}$ is mainly a function of location and secondary user transmission characteristics. FCC defines various rules for secondary TV band devices (or TVBDs, subsequently) to protect primary receivers \cite{Std:FCC10,Std:FCC12} that affect the probability of channel availability. A brief summary of the relevant FCC regulations follow:
\begin{itemize}
	\item \textbf{Permissible Channels}: A fixed TVBD may operate on any channel in $\{2:51\}\setminus\{3,4,37\}$ subject to conditions below \cite{Std:FCC10,Std:FCC12}. Further, personal/portable devices may only transmit on available channels above channel 20, $\{21:51\}\setminus\{37\}$ subject to following requirements.
	\item \textbf{Power limit}: For fixed TVBD, the maximum power delivered to antenna may not exceed 1 watt in 6 MHz with a maximum of 6-dBi gain for antenna (maximum 36-dBm of EIRP\footnote{Effective Isotropic Radiated Power}). For personal/portable TVBD, the maximum EIRP shall not exceed 20dBm per 6 MHz. If portable TVBD is transmitting in an adjacent channel to a primary transmitter, then maximum EIRP is limited to 16-dBm.
	\item \textbf{Antenna height}: The transmit antenna for fixed devices may not be more than 30 meters above the ground. In addition, fixed devices may not be located at sites where the antenna height above average terrain is more than 250 meters. Portable device antenna is assumed to be less than 3 meters above the ground.
	\item \textbf{Interference protection}: TVBD must protect digital and analog TV services within the contours defined in Table \ref{tb:table1} for various types of TV services. Fixed and portable TVBD are not allowed to transmit within a minimum separation distance from the border of protected contour that is defined in \cite{Std:FCC12} based on secondary transmitter class and height above average terrain (HAAT). Fixed devices must be outside protection regions of co-channel and adjacent channel stations. Portable devices are allowed to transmit within adjacent channel contours with a maximum power of 40 mW.
	\item \textbf{PLMRS/CMRS}: TVBD may not operate at distances less than 134-km for co-channel operation and 131-km for adjacent channel operation from metropolitan areas.
	\item \textbf{Radio Astronomy Sites}: TVBDs are not allowed to operate within 2.4 km from registered radio astronomy sites in FCC database.
	\item \textbf{Microphone Reserved}: TVBDs are not permitted to operate on the first channel on each side of TV channel 37.
\end{itemize}

The aggregate effect of FCC rules is modeled through defining protection regions for primary users, as discussed below, where no secondary is allowed to transmit. Note that protection region includes protection contour as well as minimum separation distance. Therefore, channel availability probability represents the ratio of the area where channel is considered free (according to regulations above ) to the total area of discussion.

\subsection{Primary Protection Region}
Considering a permissible WS channel $c_i$ at any location and an area $\mathcal{A}$ with $N$ co-channel secondary and $M$ adjacent-channel (either channels $c_i-1$ or $c_i+1$) primary users. For every licensed device, a protected contour \cite{Std:FCC10}, defined by FCC for different types of stations, is considered that represents the coverage area of that transmitter. This grade B contour for TV broadcasters, is a function of following parameters:
\begin{itemize}
	\item $P_t$: Primary transmitter effective power (EIRP)
	\item $\Delta$: Minimum required signal for primary receiver, defined in Table \ref{tb:table1}
	\item $f$: Frequency
	\item $h_t$/$h_r$: Tx/Rx Antenna height above average terrain (HAAT)
	\item $\Delta H$: Terrain irregularity parameter which distinguishes plains versus mountains.
	\item Service type: FCC regulates different rules for various services in TV band, such as PLMRS/CMRS\footnote{Personal Land/Commercial Mobile Radio Services} versus low power auxiliary services including wireless microphones.
	\item Environmental effects that changes propagation path loss, such as radio climate, conductivity of ground, surface refractivity, ... .
\end{itemize}

The overall protection region is an area of radius $r_p=r_{PC}+d_{MS}$ where $r_{PC}$ is the protected contour radius and $d_{MS}$ is an additional minimum separation distance, Fig. \ref{fig:coverage}. Detailed description of protection region calculation is provided in Appendix \ref{app:pr}. We define this area as the primary network cell shown in Fig. \ref{fig:cellStruct}. The secondary cells can be exist in any region beyond these primary protected cells. Therefore, the channel availability probability is defined, using an area average, as
\begin{align}
p(\Gamma) = 1 - \frac{\sum_{j=1}^{N}{\mathcal{A}_{p,co}(j)} + \sum_{k=1}^{M}{\mathcal{A}_{p,adj}(k)} -  \sum_{i}{\sum_{j}{\mathcal{A}_p(i,j)}} }{\mathcal{A}}
\label{eq:CAP}
\end{align}
where $\mathcal{A}_{p,co}(j)$ is the co-channel protection area for transmitter $j\in\{1:N\}$ and $\mathcal{A}_{p,adj}(k)$ is adjacent-channel protection area for transmitter $k\in\{1:M\}$. In most cases, as our simulation reveals, there are significant overlaps between co-channel and adjacent channel areas which is considered in $\mathcal{A}_p(i,j)$. Dependency of $\mathcal{A}_{p,co}$, $\mathcal{A}_{p,adj}$ on $\Gamma$ is removed for notational simplicity.

\begin{figure}[!t]%
	\centering
	\includegraphics[width=\figwidth]{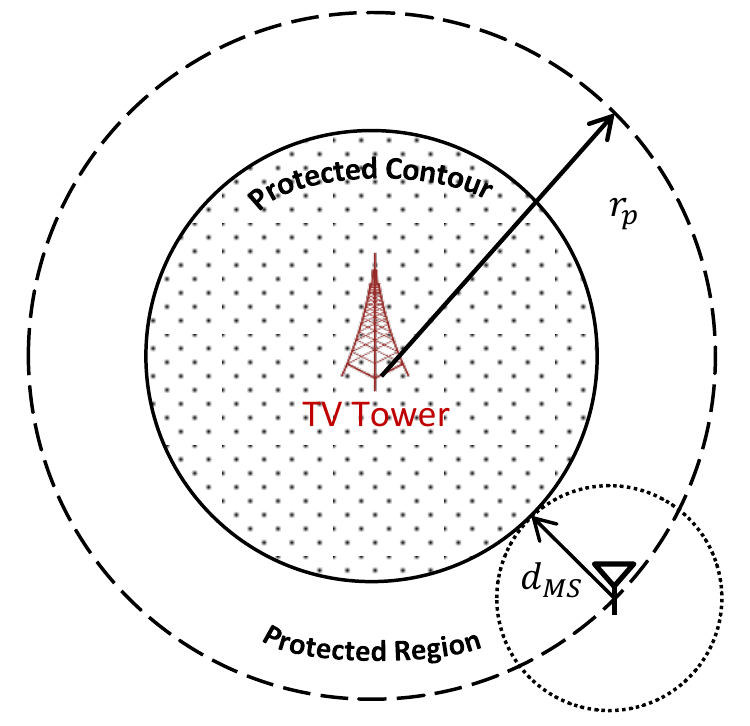}%
	\caption{Protected contour v.s. protection region defined by FCC}%
	\label{fig:coverage}%
\end{figure}

 We remark that the protection region in reality is generally an irregular area that is non-circular, due largely to varying terrain heights as a function of azimuth angles as seen at a  primary transmitter location. The FCC  simplifies this by averaging the HAAT over all azimuth into a single quantity and representing the area by a circle of radius $r_p$. This approach is acceptable for a 1st-order model for primary coverage, but not for accuracy in estimating WS network capacity. In our simulation, we will thus consider non-circular protection regions that preserves the variation of coverage distance as a function of angle.

\section{Path Loss Model}

Choosing an appropriate path loss model is very important in TVWS capacity analysis because it directly affects all subsequent results and choice of secondary parameters. There are various path loss models in the literature that has been adopted for different applications, frequency range and environments \cite{Rp:COST213,paper:hata,recom:itu,NBS:2TransLoss,ITM:Model,ITS:4Model}. Except for free space model that is derived from pure theory, most useful path loss models are based on experimental measurements; this includes
the well-known {\em HATA} model family for different terrain categories such as urban, suburban and rural areas \cite{paper:hata} that has been widely used in cellular network planning. However, such  path loss models are significantly limited in terms of accuracy in their range of parameters such as transmitter/receiver antenna height, coverage distance and frequency. Therefore, for TV tower specifications including very high altitude (as high as 700 meters) and broad coverage area (upto 100 km), more general models are required that incorporate real terrain information. We thus settle on the  Longley-Rice model that is measurement-driven and covers a wide range of input parameters, appropriate for TV coverage estimation \cite{FCC:LongleyRice}. A computer implementation of this model is provided by \cite{ITM:Model} called ITM, and is described in the next section.

\subsection{Irregular Terrain Methodology (ITM)}
ITM estimates radio propagation losses for frequencies between 20-MHz and 20-GHz as a function of distance and the variability of the signal in time and space. It is an improved version of the Longley-Rice Model \cite{EESA:1LongleyRice}, which gives an algorithm developed for computer applications. The model is based on electromagnetic theory and signal loss variability expressions derived from extensive sets of measurements \cite{NBS:2TransLoss,NTIA:3ITS}. It is applicable to point-to-area calculations with {\em point} being the location of a broadcast station or a base station for mobile service and {\em area} refers to locations of broadcast receivers or mobile stations. The area is described by the terrain roughness factor (irregularity parameter) $\Delta h$, which is defined as the interdecile value computed from the range of all terrain elevations for the area, calculated separately in every direction.

Based on Longley-Rice methodology, calculation of coverage is as follows. For analog TV, computation are made inside the conventional Grade B contour defined in Section 73.683 of the FCC rules, with the exception that the defining field for UHF channels is modified by subtracting a dipole factor. The same adjustment is needed for digital TV calculations. Modified signal strength tables for analog and digital TV are shown in Table \ref{tb:AnalogThreshold}.

\subsubsection{ITM - Input Parameters}
The following input parameters are required for a proper description of the communication link. The main parameters affecting this model are:
\begin{itemize}
	\item $f$: frequency, 20 MHz to 20 GHz.
	\item $d$: Distance between the two terminals, 1 km to 2000 km.
	\item $h_{g1},h_{g2}$: Antenna structural heights, 0.5 m to 3000 m.
	\item $pol$: Horizontal or Vertical polarization.
	\item $\Delta h$: Terrain irregularity parameter. This is the main parameter that captures the effect of terrain elevation on loss calculation.
\end{itemize}
Note that antenna heights are calculated relative to average terrain height surrounding TV transmitter. This in fact changes in different angles and results in non-circular path loss patterns just as terrain roughness parameter $\Delta h$ creates angle dependency.

\section{Interference Model}
\label{sec:InterferenceModel}
SINR is by definition the ratio of received power at receiver location to noise and interference:
\begin{align}
\text{SINR}(Q_R)=\frac{P_{sec,RX}(Q_R)}{N_0W_0+I(Q_R)}
\label{eq:SINRDef}
\end{align}
There are two sources of interference in TV white space network:
\subsection{Primary-to-Secondary} 
\begin{figure}[!t]%
	\centering
	\includegraphics[width=\figwidth]{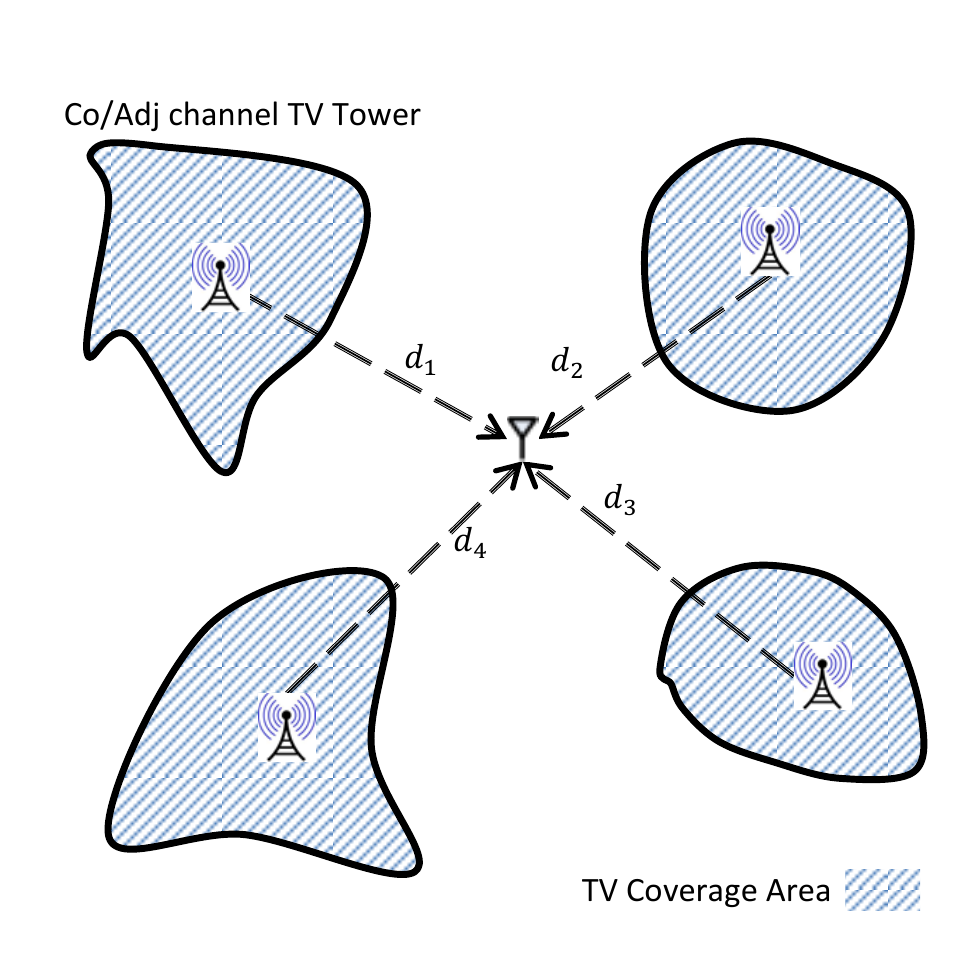}%
	\caption{Primary-to-Secondary interference}%
	\label{fig:intfrTV}%
\end{figure}

Although primary users are protected from undesired interferences from unlicensed devices, they introduce a significant source of interference to secondary users working in the same or adjacent bands. The level of interference at every location depends on the distance from the secondary receiver to nearby TV broadcasters, as shown in Fig. \ref{fig:intfrTV}, as well as other physical parameters such as antenna height.

TV transmitters introduce both co-channel and adj-channel interference to secondary users, due to signal leakage from each channel to lower/upper bands. The aggregate interference for the receiver location $Q_R$ is:
\begin{align}
I_{P2S}(Q_R) = \sum_{i=1}^{N}{(1-\eta)\frac{P_iG_i}{L_{TV}(d_i)}G_r} + \sum_{j=1}^{M}{\eta\frac{P_jG_j}{L_{TV}(d_j)}G_r}
\label{eq:TVinterf}
\end{align}
where $P_i/G_i$ is the transmitter power and antenna gain, $L_{TV}(.)$ is path loss function from TV broadcaster, $G_r$ is the receiver antenna gain and $N/M$ is number of co-channel/adj-channel surrounding TV towers (we consider TV transmitters up to a distance of 300 km in numerical calculation section). $\eta$ is the leakage factor for TV transmitters defined as the ratio of power transmitted in upper/lower band to total power and $d_i$ is distance to primary transmitter $i$. Note that, location dependency, $I_{P2S}(Q_R)$ is hidden inside distance factor $d_i$.

\begin{figure}%
\centering
\includegraphics[width=\figwidth]{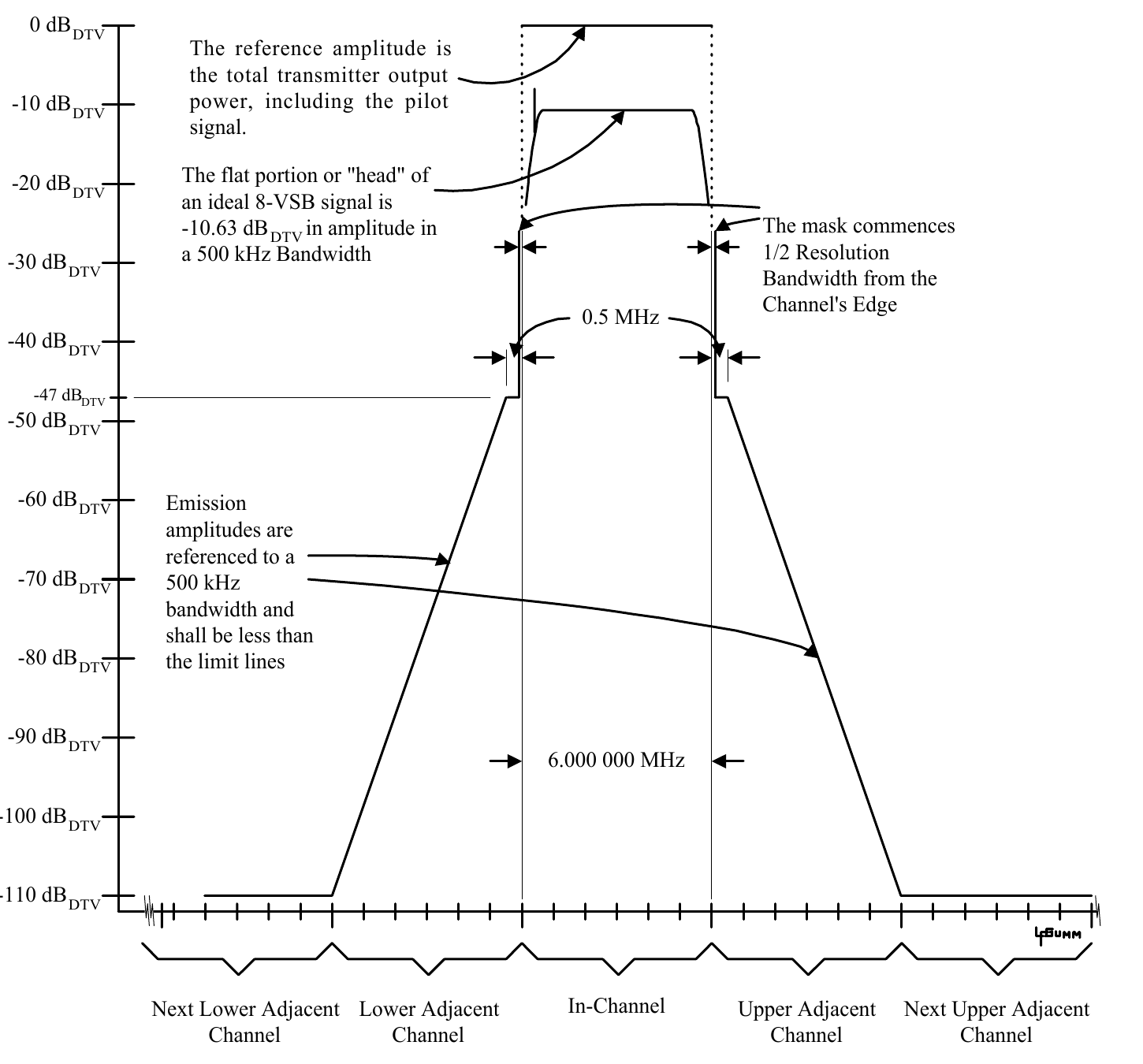}%
\caption{8-VSB Full Service Transmitter Emission Limits \cite{Std:IEEEP1631D3}}%
\label{fig:TxMask}%
\end{figure}

In order to consider adjacent channel interference in simulation, we consider a practical transmission mask, introduced for 8-VSB standard \cite{Std:TxMaskIEEE}, shown in Fig. \ref{fig:TxMask} for full service digital TV transmitters. It defines maximum power leakage to upper and lower channels for a maximum of two channel distance. Similar masks are defined for TV translators and low power TV broadcasters \cite{Std:TxMaskIEEE}. Full service transmitter mask is defined as below:
\begin{itemize}
	\item In the range between Channel Edge and 500 kHz from the Channel Edge: $Emission \leq -47 dB_{DTV}$\footnote{$dB_{DTV}$ is the relative power with respect to \em{total} power in the transmitter's 6 MHz Channel including the pilot}
	\item More than 6 MHz from Channel Edge: $Emission \leq -110 dB_{DTV}$
	\item At any frequency between 500kHz and 6 MHz from the Channel Edge: $Emission \leq -(11.5(|\Delta f|-0.5)+47) dB_{DTV}$, with $\Delta f$ being the frequency difference in MHz from the Channel Edge.
\end{itemize}
By defining the emission mask function $E(f)$ as above for full service DTV, the overall leakage to adjacent channels with respect to total power is:
\begin{align}
\eta_{\pm1}(DTV) &= \int_{0}^{6}{10^{E(f)/10}df}=1.75*10^{-5} \nonumber \\
\eta_{\pm2}(DTV) &= \int_{0}^{6}{10^{-110/10}df}\approx 0\nonumber
\end{align}
Following the same procedure for LPTV DTV or translator services results in:
\begin{align}
\eta_{\pm1}(LPTV) &= \int_{0.5}^{3}{10^{-(1.15(f-0.5)+4.7)}df} \nonumber \\
  								&+ 5*10^{-5.7}+3*10^{-7.6} = 1.76*10^{-5} \nonumber \\
\eta_{\pm2}(LPTV) &= \int_{0}^{6}{10^{-76/10}df} = 1.51*10^{-7}\nonumber
\end{align}

\subsection{Secondary-to-Secondary} The interference that is experienced by a white space receiver from {\em other } TVBD users has severe impacts on its performance. As discussed before, the area $\mathcal{A}$ is divided to multiple cells and each channel is used several times in non-adjacent cells. The minimum distance that allows the same frequency to be reused will depend on many factors, such as the number of co-channel cells in the vicinity of the center cell, the type of geographic terrain contour, the antenna height, and the transmitted power at each cell site, \cite{book:WilliamLee}.

The frequency reuse distance D can be determined from
\begin{align}
D = \sqrt{3K}r_{cell}
\end{align}
Where $K$ is the frequency reuse pattern, defined by shift parameters $K=i^2+ij+j^2$. In theory, increasing $D$ will reduce the chance of co-channel interference and is desired. On the other hand, spectrum inefficiency will also increase as the ratio of $q=\frac{D}{r_{cell}}$ increases. The goal is to obtain the smallest $K$ that maximizes efficiency and still meets the protection requirements on incumbents. This involves estimating co-channel interference and selecting the minimum frequency reuse distance $D$ feasible.

Here, we assume that secondary cell size is fixed and is determined by the coverage area corresponding to the secondary transmit power \cite{book:WilliamLee}. For a homogeneous secondary network with fixed cell size, the co-channel interference is {\em independent} of the transmitted power of each cell, i.e. the receiver threshold at a mobile secondary receiver is adjusted to the size of the cell. The received carrier-to-interference ratio at the desired mobile receiver is \cite{book:WilliamLee}
\begin{align}
\frac{C}{I}=\frac{C}{K_I\sum_{i,j\in[0,1,...]}{I_{i,j}}}
\end{align}
where $K_I$ is the number of interferer at each tier and $I_{i,j}$ is the interference of the co-channel cell shifted by $i, j$. In a fully equipped hexagonal-shaped cellular system, $K_I=6$ and $I_{i,j}=P_{sec,TX}G_{sec}^2L_{sec}(D_{i,j})$ where $D_{i,j}$ is the distance from $(i, j)$'s interfering cell and $D_{i,j}=\sqrt{i^2+ij+j^2}D$. The overall experienced secondary-to-secondary interference is
\begin{align}
I_{S2S} = P_{sec,TX}G_{sec}^2K_I & \sum_{i=0}^{\infty}{\sum_{j=i,j\not =0}^{\infty}{L_{sec}(D_{i,j})}}
\end{align}
In practice, the closest cells are the prominent interferers and the sum above is practically limited to $i,j=2$ instead of infinity. Note that in theory, $I_{S2S}$ depends on precise location of the receiver inside the cell but we calculate the interference for the center of the cell site. Therefore, this will relax dependency of $I_{S2S}$ on $Q_R$. The resulting SINR is
\begin{align}
\text{SINR}(Q_R)&=\frac{P_{sec}/L_{sec}(r_{cell})}{N_0W_0+ I_{P2S}(Q_R) + I_{S2S}}
\end{align}
Using (\ref{eq:AvgChan}), (\ref{eq:CAP}) and (\ref{eq:SINRDef}), the average capacity per cell for each individual TVWS channel is
\begin{align}
\overline{C_{cell}}(\Gamma) &= \frac{p(\Gamma)}{K}W_0 \int_{Q_R}{\log_2\left(1+\frac{P_{sec}/L_{sec}}{N_0W_0 + I_{P2S}(Q_R)+I_{S2S}}\right)dQ_R}
\label{eq:AvgChanwithSINR2}
\end{align}

\section{Numerical Calculations}
\begin{figure}[!t]%
	\centering
	\includegraphics[width=\figwidth]{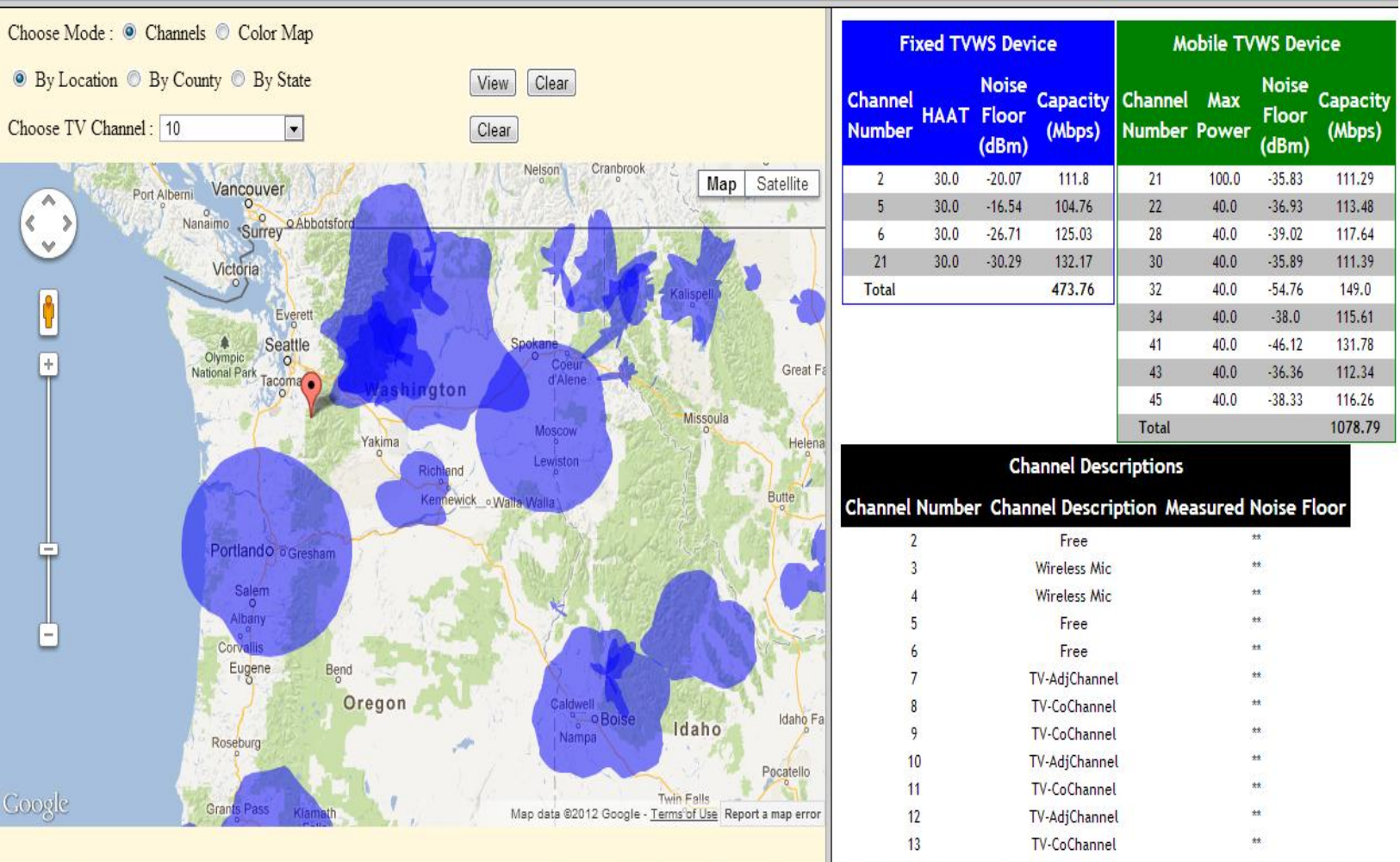}%
	\caption{TVWS simulation engine, developed at University of Washington, a cloud based simulator for TVWS. Protection regions are shown for channel 10 across Washington state; List of available channels for fixed and mobile devices, as well as estimated interference level and capacity are shown for available channels \cite{ele:SpecObs}.}%
	\label{fig:SpecObs}%
\end{figure}

In this section, numerical results are provided by evaluation of the analysis in previous sections. The main focus is to explore dependency of network capacity on various parameters and subsequent optimization of parameter choices. All simulations are performed on a TVWS simulation platform developed at University of Washington, a snapshot of which is shown in Fig. \ref{fig:SpecObs}. This platform models protection regions for all primary transmitters registered in the FCC database \cite{FCC:techrep1}, and applies all FCC regulations to determine available channels to secondary devices. It also estimates interference level and capacity at any location inside the United States. Significantly, actual terrain information\footnote{Terrain data, obtained from Globe \cite{site:globe},  used in calculation of two parameters, first, transmitter HAAT, second, terrain irregularity $\Delta H$, both versus angular direction with 1-degree resolution.} was used for accurate computation of path loss function.

\begin{figure}[!t]%
	\centering
	\includegraphics[width=\figwidth]{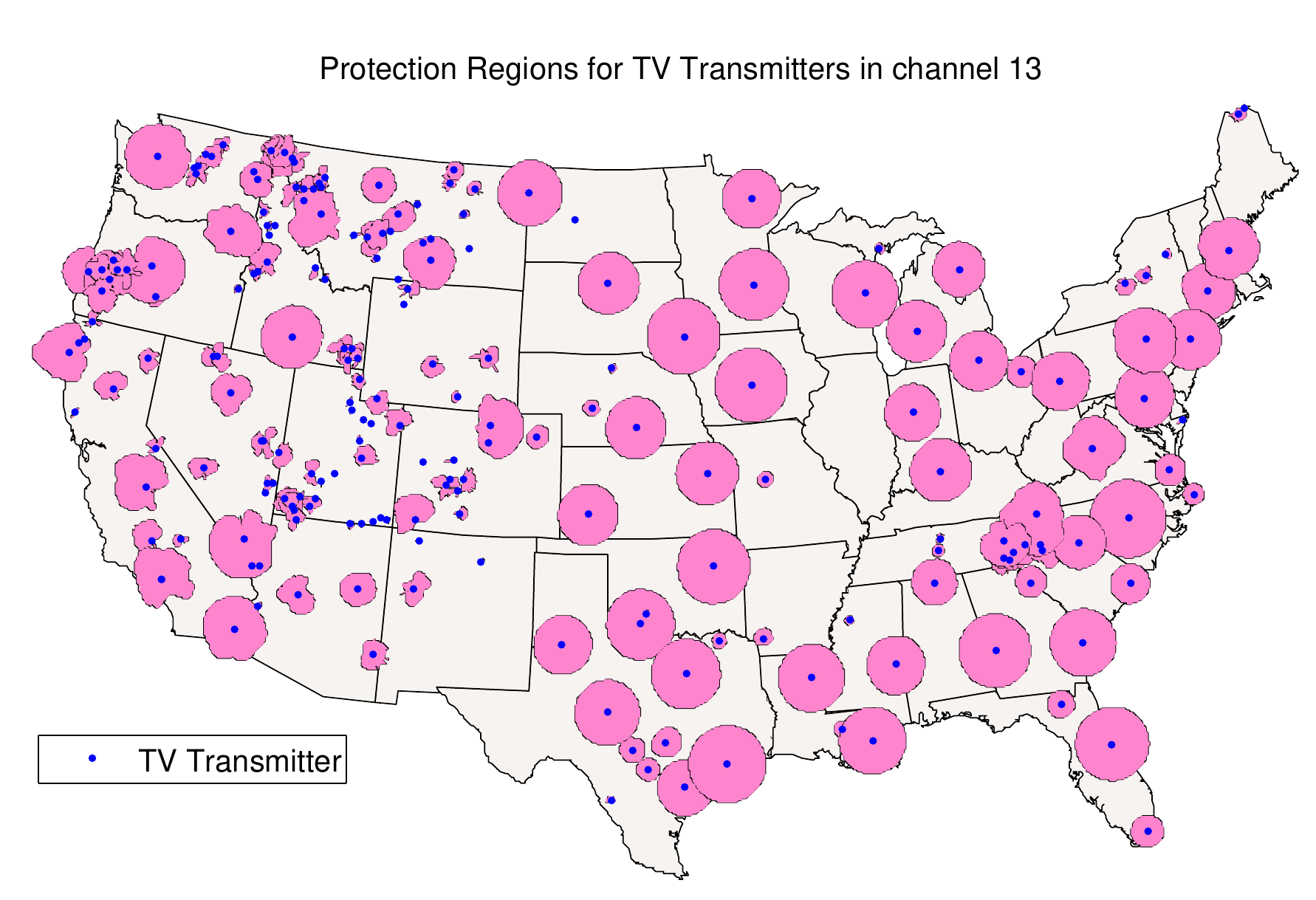}%
	\caption{Protection region for TV transmitters broadcasting on channel 13; Using Longley Rice model in area mode}%
	\label{fig:Ch13Pr}%
\end{figure}

\begin{figure}[!t]%
	\centering
	\includegraphics[width=\figwidth]{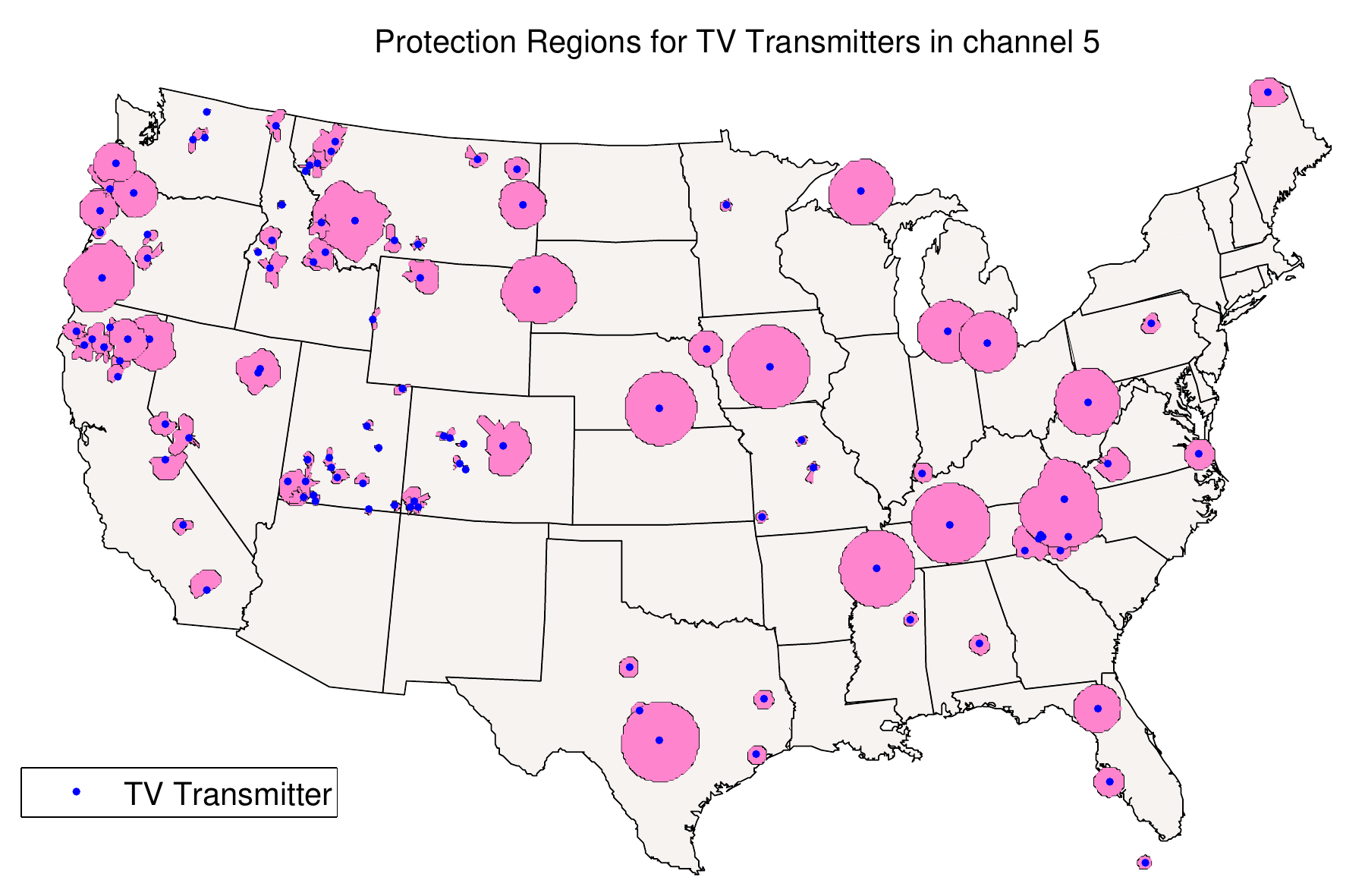}%
	\caption{Protection region for TV transmitters broadcasting on channel 5; Using Longley Rice model in area mode}%
	\label{fig:Ch5Pr}%
\end{figure}

Fig. \ref{fig:Ch13Pr} and \ref{fig:Ch5Pr} shows the protection regions for TV transmitters on channel 13 and 5 across the continental US, using Longley-Rice methodology in area mode for path loss prediction. The spatial distribution of transmitters as well as non-circularity of protections regions, as a result of terrain variations, are well highlighted in the figures. Note the general lack of structure in the distribution of TV transmitters for channels 5, 13, indicating the lack of any prior planning (as in cellular layout) for TV broadcast.


\subsection{Channel Availability Statistics}

FCC regulation restricts TVBD in many aspects from maximum power and antenna height to not using adjacent channel to active TV stations. In this section, we provide some {\em statistical information} about how these rules affects the total number of available WS channels for {\em fixed} and {\em portable} devices.

Table \ref{tb:chanAvg} shows the average number of available channels for {\em fixed} and {\em portable} TVBD in various frequency bands. In addition, the number of channels that are used by TV transmitters (Busy channels) as well as the total number of channels that are not released for unlicensed operation (because they are adjacent to busy channels or reserved for wireless microphone) are provided. At the bottom of the table is the final utilization factor that is achieved by permitting unlicensed operation and it is defined as $CUF=1-\frac{\mbox{Unused Channels}}{\mbox{Total TV Channels}}$.
\begin{table}[tbp]
\caption{Average number of available channels for different TVBD classes. Average numbers are provided for the entire United States for 10000 uniformly distributed locations.}
\label{tb:chanAvg}
\renewcommand{\arraystretch}{1.3}
\begin{tabular}{ l l l l l }
\toprule
Device Type        				& LVHF (2:6)  & HVHF (7:13) & LUHF (14:51\textbackslash37) & Total \\ \hline
Total Available    				& 2.36				& 2.59        & 21.77        & 26.73\\
Fixed Devices      				& 2.36        & 2.59        & 15.2         & 20.17 \\
Portable/Personal  				& 0           & 0           & 18.79        & 18.79 \\
Microphone Reserved				& 2           & 0           & 2            & 4 \\
Busy Channels by TV				& 0.45        & 2.22        & 10.43        & 13.11 \\
Unused Channels    				& 2.18        & 2.19        & 4.67         & 9.05 \\ \hline
Channel Utilization Factor& 56\%				& 68.7\%				& 87.4\%  			 & 81.5\% \\
\bottomrule
\end{tabular}
\end{table}
As the table suggests, an average of 9.09 channels are still left unused even with operation of TVWS devices, which is $\approx 19\%$ of all channels. By repeating this simulation for non-uniformly distributed locations, selected separately from urban and rural areas, interesting results are observed. As shown in Table \ref{tb:chanAvgUrban} and Table \ref{tb:chanAvgRural}, the number of available WS channels is highly dependent on population density. The total number of channels in rural areas are twice as many as in the urban areas, mainly because of significant reduction in the number of active TV transmitters; however, the number of unused channels is approximately the same in both.
\begin{table}[tbp]
\caption{Average number of available channels for Urban areas; A minimum population density of 1000 person per Sq. miles is considered as urban area.}
\label{tb:chanAvgUrban}
\renewcommand{\arraystretch}{1.3}
\begin{tabular}{ l l l l l }
\toprule
Device Type        & LVHF (2:6)  & HVHF (7:13) & LUHF (14:51\textbackslash37) & Total \\ \hline
Total Available    & 1.69				 & 0.76				 & 7.85					& 10.3\\
Fixed Devices      & 1.69 			 & 0.76				 & 1.95 				& 4.4\\
Portable/Personal  & 0					 & 0					 & 7.3					& 7.3\\
Microphone Reserved& 1.81				 & 0					 & 1.88					& 3.69\\
Busy Channels by TV& 0.91				 & 3.92			   & 23.2				  & 28\\
Unused Channels    & 2.4				 & 2.32				 & 4.67				  & 9.4\\\hline
Channel Utilization Factor& 52\% & 67\%				 & 87\%					& 81\%\\
\bottomrule
\end{tabular}
\end{table}

\begin{table}[tbp]
\caption{Average number of available channels for Rural areas; With population density less than 1000 person per Sq. miles}
\label{tb:chanAvgRural}
\renewcommand{\arraystretch}{1.3}
\begin{tabular}{ l l l l l }
\toprule
Device Type        & LVHF (2:6)  & HVHF (7:13) & LUHF (14:51\textbackslash37) & Total \\ \hline
Total Available    &2.20				 & 1.60  			 & 17.4				& 21.2\\
Fixed Devices      &2.2 				 & 1.60				 & 8.82				& 12.63\\
Portable/Personal  &0						 & 0					 & 15.56		  & 15.56\\
Microphone Reserved&1.87				 & 0					 & 1.98				& 3.85\\
Busy Channels by TV&0.5					 & 2.75				 &14.05				&17.3\\
Unused Channels    &2.3					 & 2.64				 &5.23				&10.17\\\hline
Channel Utilization Factor& 54\% & 62\% 			 &86\%			  &79\%\\
\bottomrule
\end{tabular}
\end{table}

\begin{table}[tbp]
\caption{Average number of available, busy, and unused channels for selected cities in United States}
\label{tb:chanAvgCities}
\renewcommand{\arraystretch}{1.3}
\begin{tabular}{ l l l l l l l}
\toprule
Location          				& New York & Houston & Chicago & Seattle & Miami & Denver\\ \hline
Total Available    				& .05 		 & 5.16 	 & 5.57    & 12.81   & 1.65  & 3.176\\
Fixed Devices      				& 0 			 & 2.13    & 1.06    &  4.67   & 0     & 1.058\\
Portable/Personal  				& .05 		 & 3.04	   & 4.50	   &  9.8    & 1.65  & 2.176\\
Microphone Reserved				& 3.93 		 & 3.93	   & 3       & 4       & 3     & 4\\
Busy Channels by TV				& 37.1   	 & 31.25   & 31.51   & 25.5    & 38.23 & 37.5\\
Unused Channels    				& 8.10     & 11.58   & 9.92    & 10.7    & 8.11  & 8.29\\ \hline
Channel Utilization Factor& 81\%     & 76\%    & 80\%    & 78\%    & 83\%  & 83\%\\
\bottomrule
\end{tabular}
\end{table}
 
\begin{figure}%
\centering
\includegraphics[width=\figwidth]{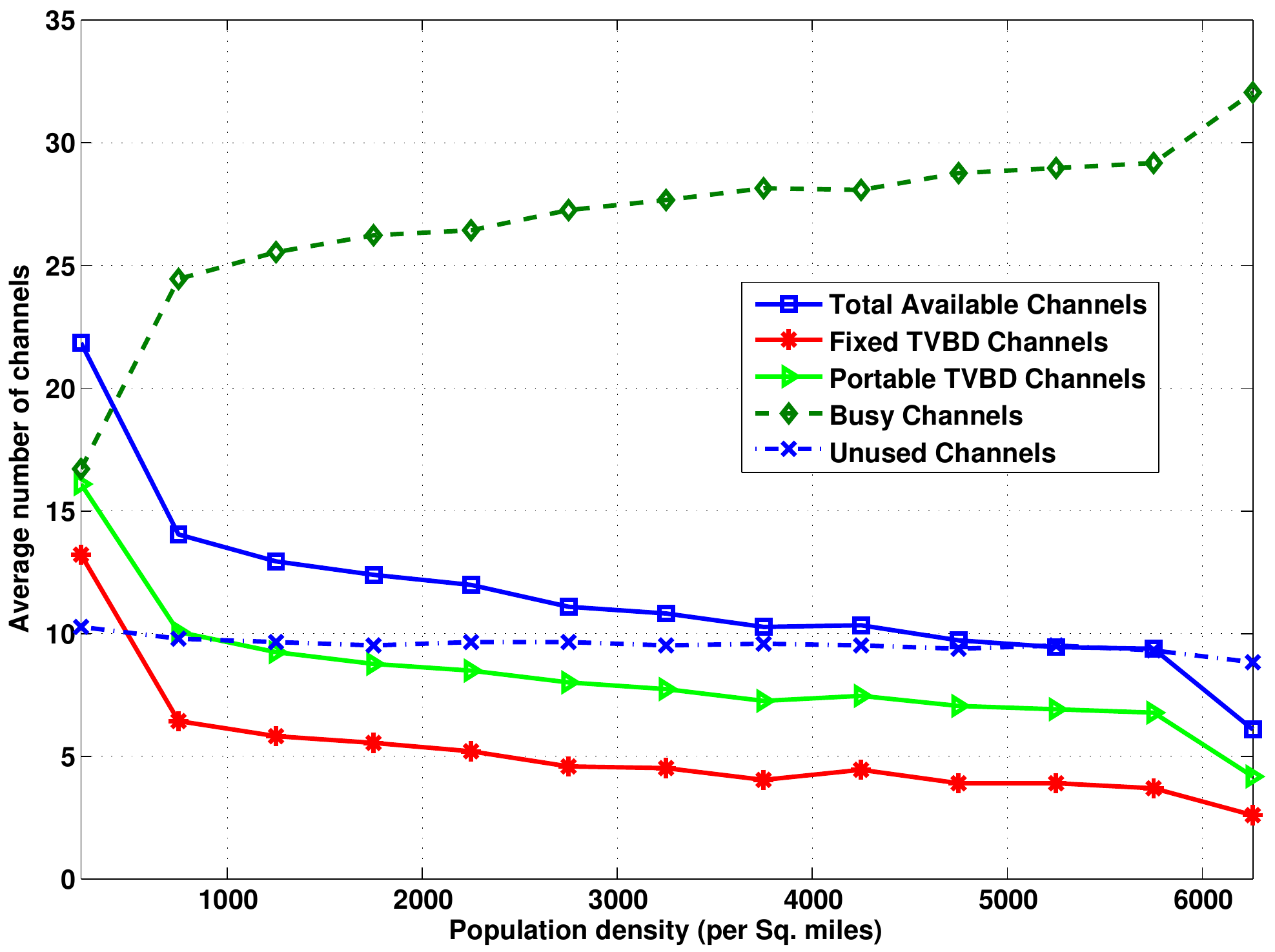}%
\caption{Average number of available channels vs. population density, calculated over the entire United States \cite{site:census}.}
\label{fig:ChanVsPopulation}%
\end{figure}

\begin{figure}%
\centering
\includegraphics[width=\figwidth]{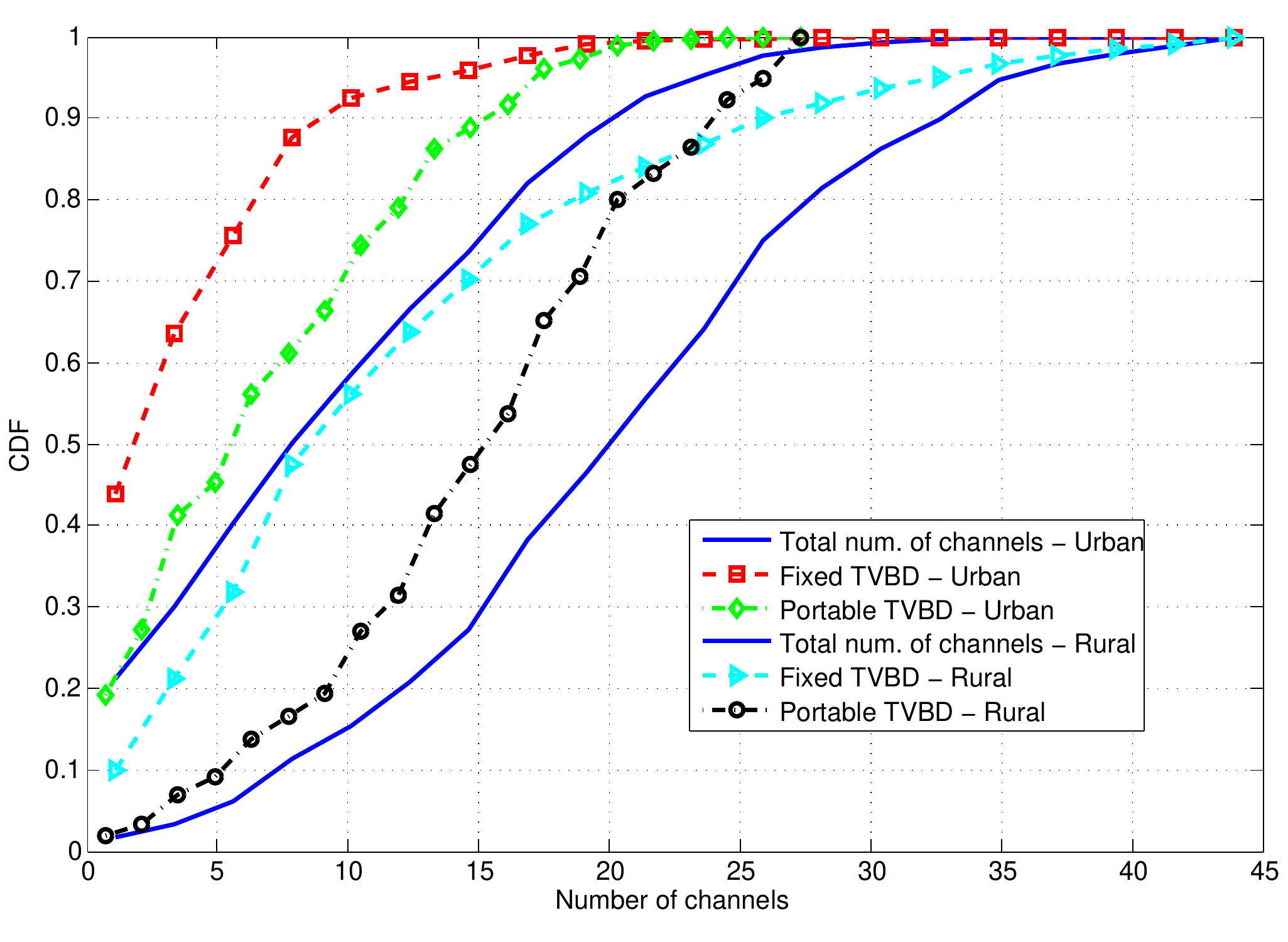}%
\caption{Cumulative density function of number of available channels for different TVBD classes. Statistics are provided for the entire United States.}%
\label{fig:PdfFreeChan}%
\end{figure}

 The variation of the average number of channels versus population density is also of interest. Fig. \ref{fig:ChanVsPopulation} shows the total number of available channels, sub-divided into those for fixed and portables devices. There is a noticeable big reduction in TVWS channels as population density approaches 1000 per sq. mile (representing transition from rural to semi-urban areas) and thereafter, the changes are slower. The actual number of available WS channels does not change much with population density, as noted in Tables \ref{tb:chanAvgUrban} and \ref{tb:chanAvgRural}.
Fig. \ref{fig:PdfFreeChan} illustrates cumulative density function (CDF) for total number of available channels as well as number of available channels for fixed and portable devices, in urban and rural areas, defined as:
\begin{align}
CDF(x) = Pr[\mbox{Number of channels }\leq x]
\end{align}
The rural CDF is shifted by approximately 10 channels, relative to urban CDF which again highlights further availability of TVWS in rural places.

\subsection{Primary to Secondary Interference}
\begin{figure}%
\centering
\includegraphics[width=\figwidth]{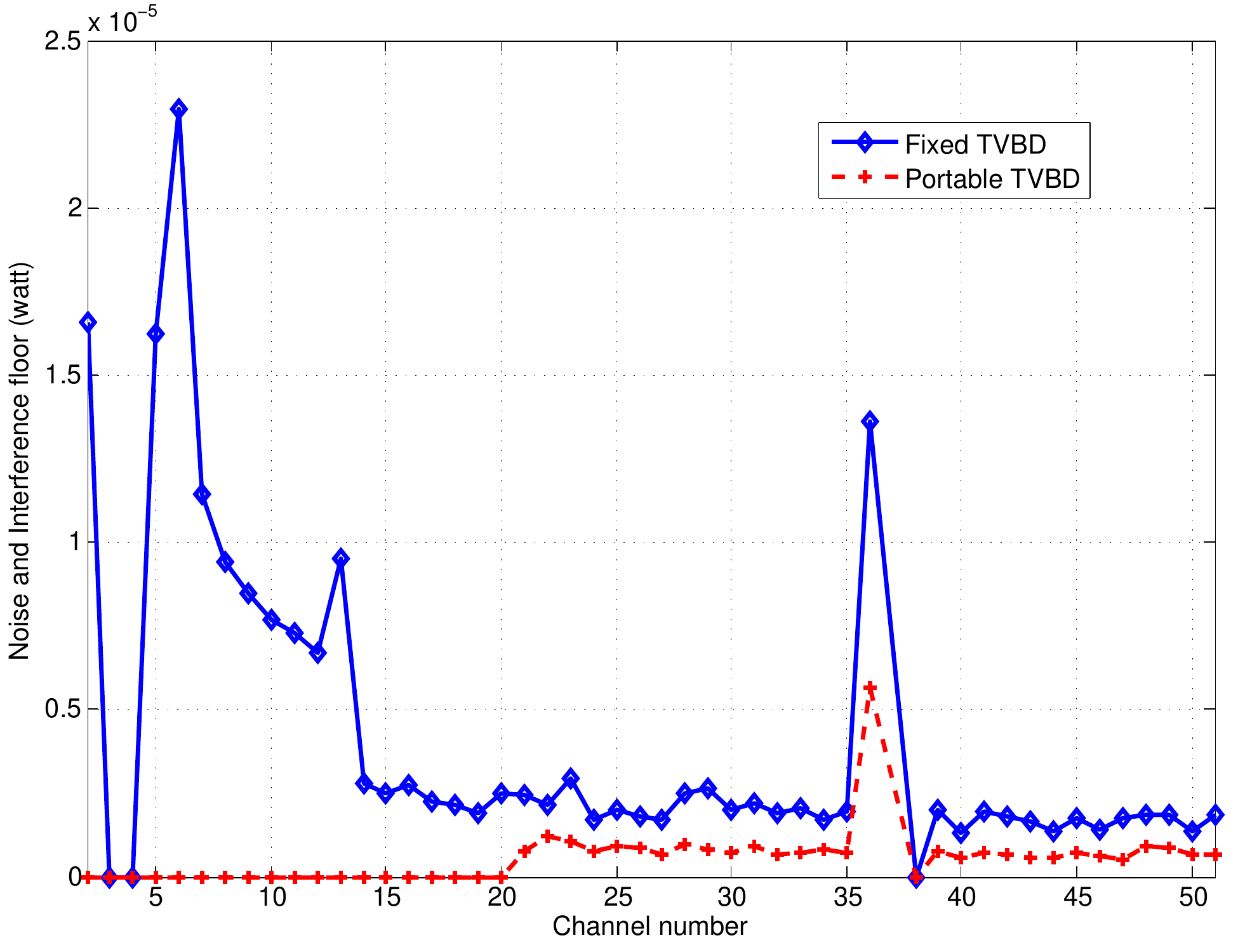}%
\caption{Noise and interference floor for {\em fixed} and {portable} TVBD, (\ref{eq:TVinterf}).}%
\label{fig:noiseInterf}%
\end{figure}

Primary transmitters are usually of very high power with a poor transmission mask, as shown in Fig. \ref{fig:TxMask}. Therefore, they have considerable out-of-band emissions up to two adjacent channels, which significantly reduces secondary user's SINR. As a result, TV {\em white space} is in fact {\em gray space} with different levels of pollution in different channels. Fig. \ref{fig:noiseInterf} illustrates noise and interference levels in all permissible channels for {\em fixed} and {\em portable} devices. The zeros in the figure correspond to channels that are not available for the associated device. For example, channel 37 is not permissible for both and channels $\{2:20\}$ are not available for {\em portable} devices.

Interestingly, the level of interference to {\em fixed} devices is much larger than {\em portables}. This is due to higher antennas that are used for {\em fixed} transmitters (up to $30$ m. is allowed according to FCC regulation\cite{Std:FCC12}) while {\em portable} transmitter antenna is supposed to be less than $3$ m. high. This will impose a trade-off between antenna height and capacity that is considered in the following section.

Fig. \ref{fig:noiseInterf} also highlights the effect of frequency on experienced interference; while channel 36 could be considered as an outlier, the interference level is inversely proportional to frequency. Therefore, at lower channels with better propagation characteristics (which is usually desired because of increased secondary link capacity and better range), secondary users receive stronger interference from nearby TV broadcaster and it degrades their capacity.

\section{Practical Trade-Offs in Secondary Network Design}

\subsection{Link Capacity vs. TVBD Power}
\begin{figure}[!t]%
	\centering
	\includegraphics[width=\figwidth]{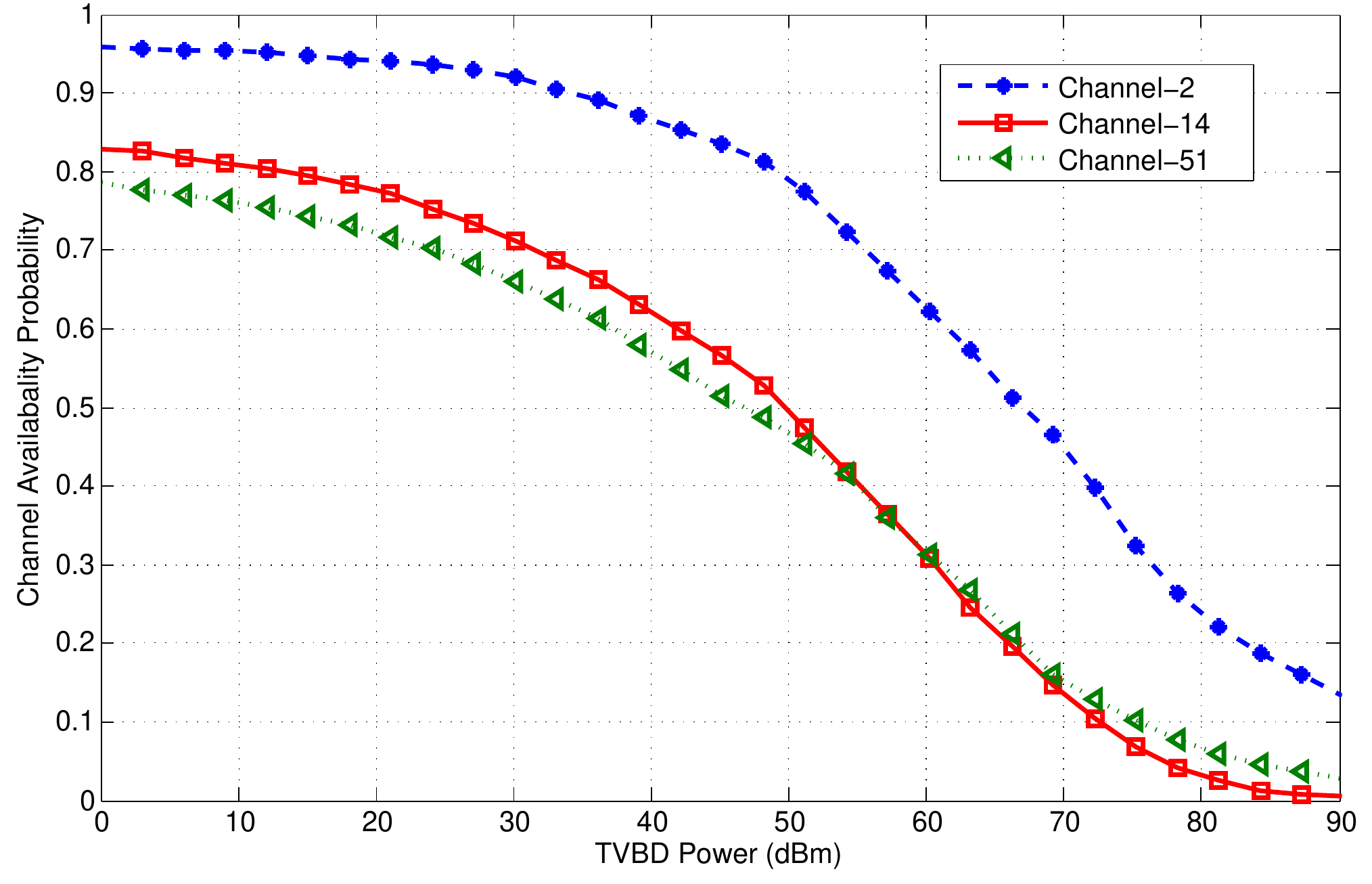}%
	\caption{Channel availability probability vs. TVBD power. Longley-Rice model is used for TV broadcasters and HATA model for 1-km distanced TVBDs. TVBD height is set to 30-m according to FCC regulations. Results are averaged over the entire United States.}%
	\label{fig:Ch13CAP}%
\end{figure}

\begin{figure}[!t]%
	\centering
	\includegraphics[width=\figwidth]{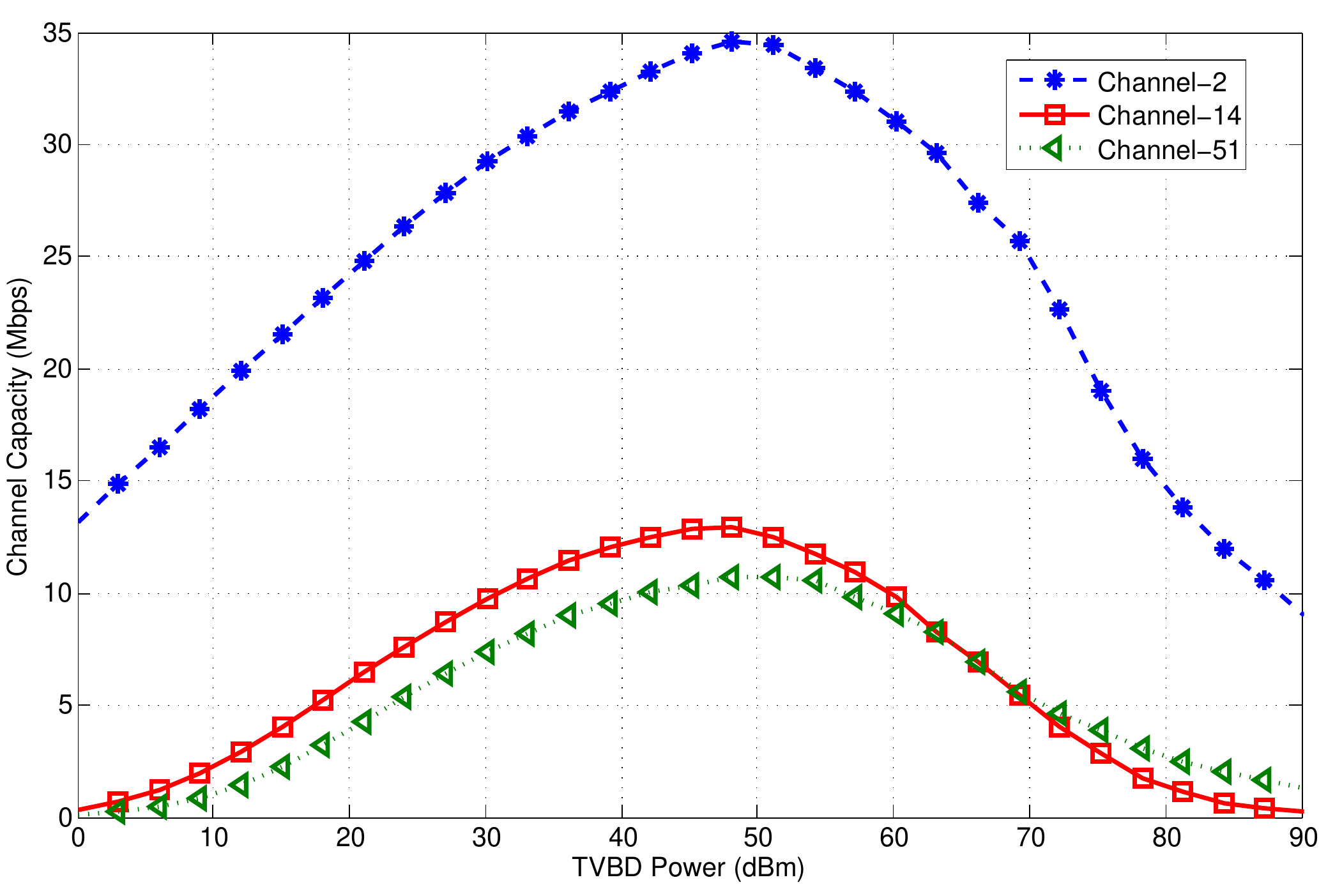}%
	\caption{Secondary network capacity vs. TVBD power. Longley-Rice model is used for TV broadcasters and HATA model for 1-km distanced TVBDs. TVBD height is set to 30-m according to FCC regulations. Eq. (\ref{eq:AvgChanwithSINR2}) is used for a pair of hypothetical secondary users exploiting the entire WS capacity in each channel. Results are averaged over the entire the entire United States.}%
	\label{fig:NetworkCapacityVsPower}%
\end{figure}

Suggested by (\ref{eq:AvgChanwithSINR2}), link capacity is a function of available bandwidth $p(\Gamma)W$ and SINR = $\frac{P_{sec}/L_{sec}}{N_0W_0+I}$. Given that the bandwidth is fixed, in order to increase link capacity in a channel, secondary power $P_{sec}$ should be increased. In contrast to regular communication systems, higher power in TVWS has a negative effect on link capacity since it expands protection region of TV transmitters. As a result, channel availability probability $p(\Gamma)$ defined in (\ref{eq:CAP}) decreases and average link capacity will diminish. The natural question is of course {\em what is the optimum secondary power?} In order to see the relative effects of secondary power, Fig. \ref{fig:Ch13CAP} shows channel availability probability $p(\Gamma)$ as a function of secondary power for three different channels. Obviously, increasing power will decrease probability of finding channel available. Channel 2 is generally more available than channel 14 and 51 since there are fewer active TV transmitters on this band.
Fig. \ref{fig:NetworkCapacityVsPower} shows average network capacity versus TVBD transmitter power. Increasing power initially increases capacity by improving SINR. However, as power increases further, the reduction in the number of locations where the channel is available (outside the protected region in Fig. \ref{fig:coverage}) leads to capacity decreasing. The interesting point is that all channels are optimized around the same secondary power even though the frequency band is considerably different (57-MHz for channel 2 and 695-MHz for channel 51) which results in smaller protection regions at higher frequency (excessive path-loss). This is mainly because there are less number of active transmitters at lower frequencies than at the higher ones which compensate for bigger protected regions.

\subsection{Link Capacity vs. TVBD Antenna Height}
\begin{figure}[!t]%
	\centering
	\includegraphics[width=\figwidth]{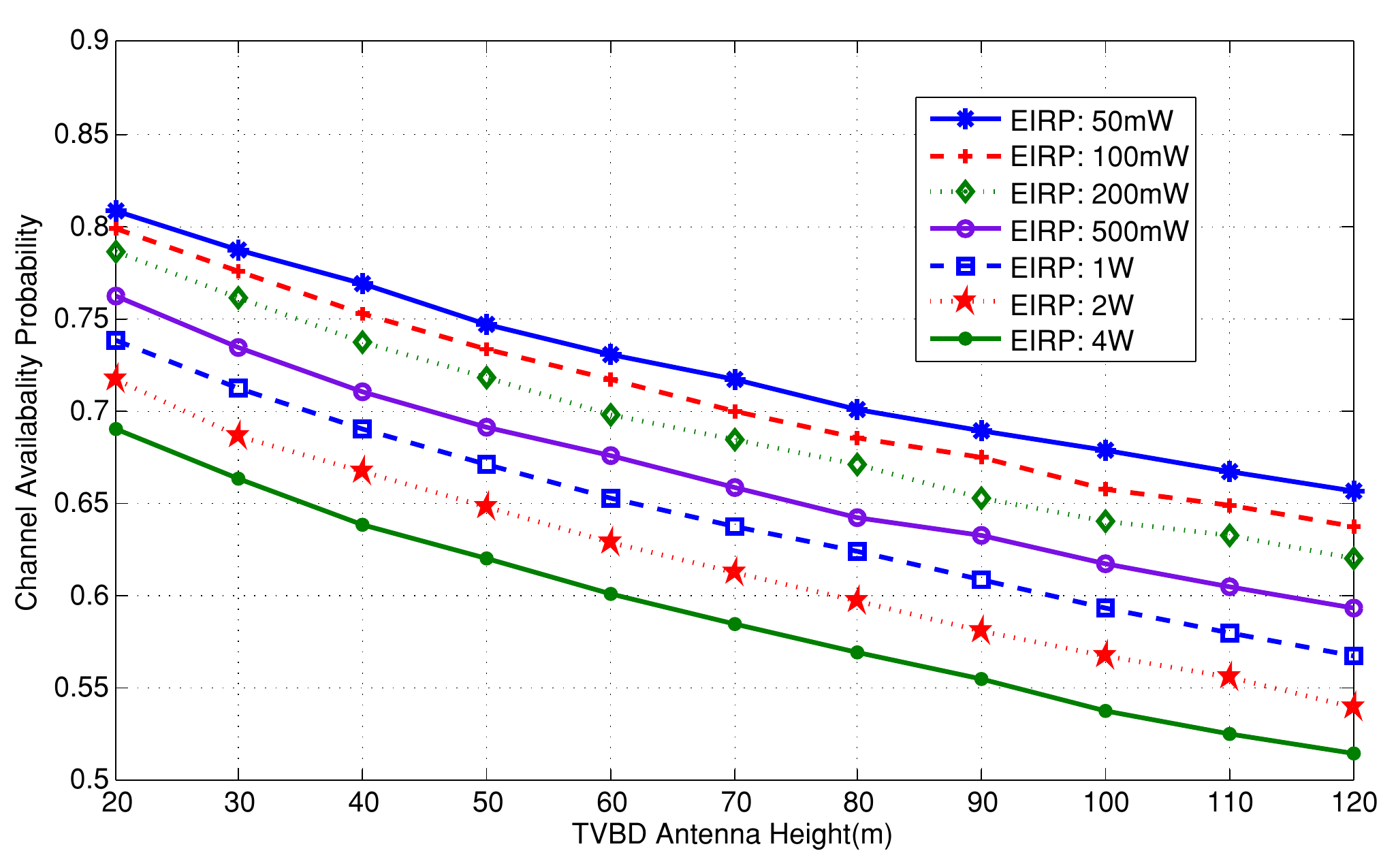}%
	\caption{Channel availability probability vs. TVBD antenna height. Longley-Rice model in area mode is used for TV transmitters and HATA model for 1-km distanced TVBDs. Results are provided for various transmission powers over channel 14. Calculation is provided for the entire United States.}%
	\label{fig:ChanProbVsAntH}%
\end{figure}

The height of the secondary user's antenna is also affecting network capacity by changing path loss for TVBD transmitter. Increasing antenna height has following effects on network capacity:
\begin{itemize}
	\item Secondary to secondary path loss decreases, higher capacity.
	\item Minimum distance increases, hence $p(\Gamma)$ reduces and results in lower capacity.
	\item Primary to secondary interference increases, lower capacity.
\end{itemize}
The overall effect of increasing antenna height depends on transmitter power, link distance and other fundamental parameters in (\ref{eq:AvgChan}). But the trade-off is obvious that increasing height will not continuously enhance capacity and there should be an cutting point. This optimization rises from the counter effects of higher received power at the receiver against less channel availability and more interference. Fig. \ref{fig:ChanProbVsAntH} shows dependency of channel availability probability on TVBD antenna height for various transmission powers. Increasing antenna height monotonically reduces channel accessibility due to extended protection regions for TV transmitters.

\begin{figure}[!t]%
	\centering
	\includegraphics[width=\figwidth]{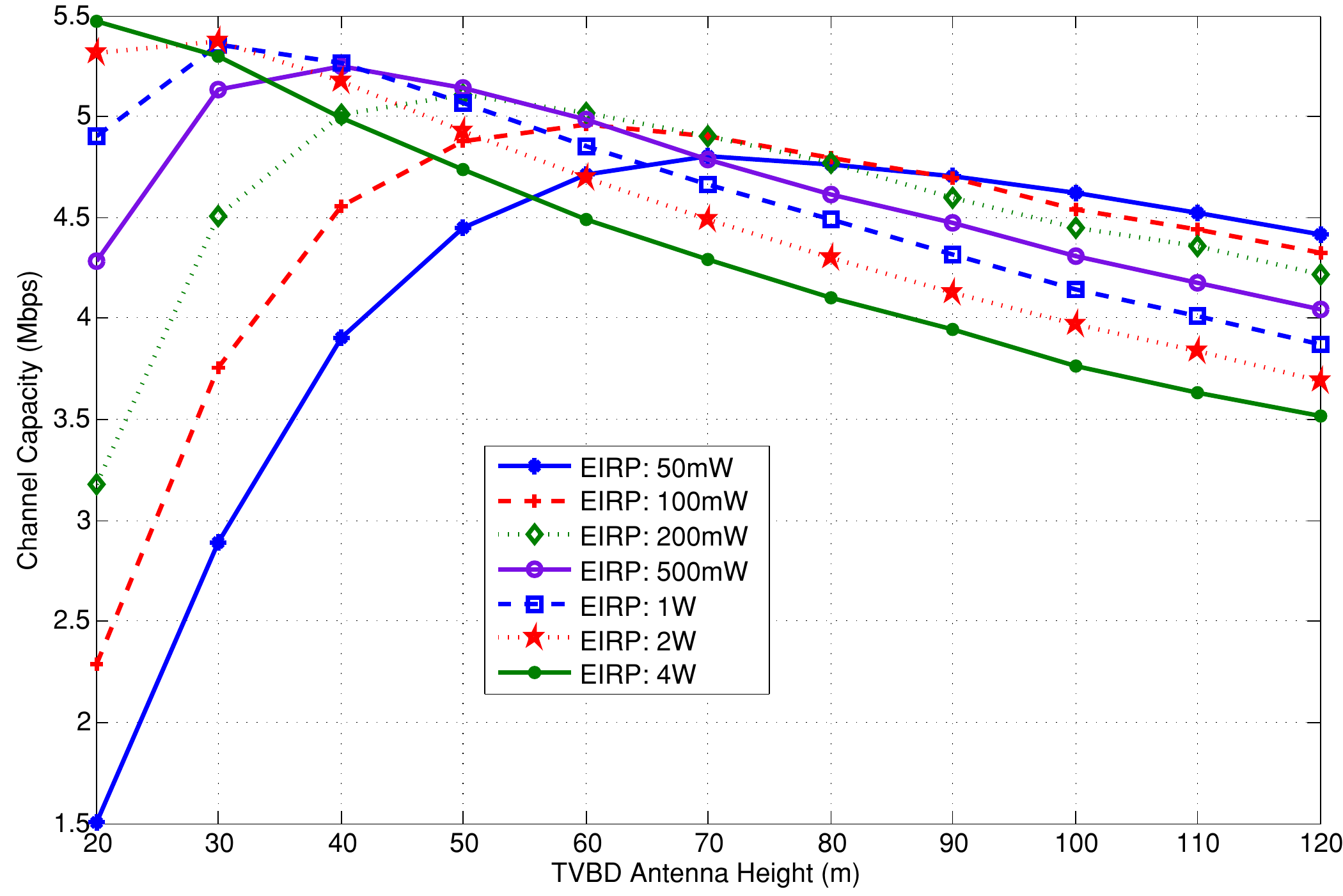}%
	\caption{Secondary network capacity vs. TVBD antenna height. HATA model is used for 1-km distanced TVBDs. Cell size is assumed to be 2-km wide. Results are provided for various transmission powers in channel 14. Average results are provided over the entire United States.}%
	\label{fig:NetworkCapacityVsAntH}%
\end{figure}

Fig. \ref{fig:NetworkCapacityVsAntH} displays secondary network capacity as a function of TVBD antenna height. For lower power scenarios (EIRP$\leq$2W), increasing height will improve capacity through a certain point where extended interference becomes dominant and capacity degrades. As can be seen from this figure, the higher the transmission power, the lower this optimum antenna which is reasonable because higher transmission power translates to amplified interference from secondary users in adjacent cells. For higher power scenarios (EIRP$\geq$4W) interference is so dominant that increasing height will always decrease capacity.

\subsection{Link capacity vs. Terrain Irregularity}
The characteristics of propagation environment is modeled through $\Delta H$ parameter in ITM model. It ranges from $\Delta H=0$ for extremely flat to $\Delta H=500$ for rugged mountainous-type area. Our previous simulations were based on calculating a $\Delta H$ in every direction around TV transmitters (resulted in non-circular contour models) for computation of protection contour. The effect of this parameter on network capacity is through $p(\Gamma)$ in (\ref{eq:CAP}) which directly modifies the overall capacity (\ref{eq:AvgChanwithSINR2}). Precise evaluation of $\Delta H$ is very critical since it significantly affects ITM path loss model (and any other empirical model). Calculating the exact value of $\Delta H$ is very challenging mainly because available terrain information is sparse, for example Globe \cite{site:globe} provides terrain height for every 30 arc-seconds (or approximately every 1 {\em km} in latitude and longitude) which may easily miss localized tall buildings or skyscrapers. Therefore, it is necessary to understand how the value of $\Delta H$ affects $p(\Gamma)$.

\begin{figure}[!t]%
	\centering
	\includegraphics[width=\figwidth]{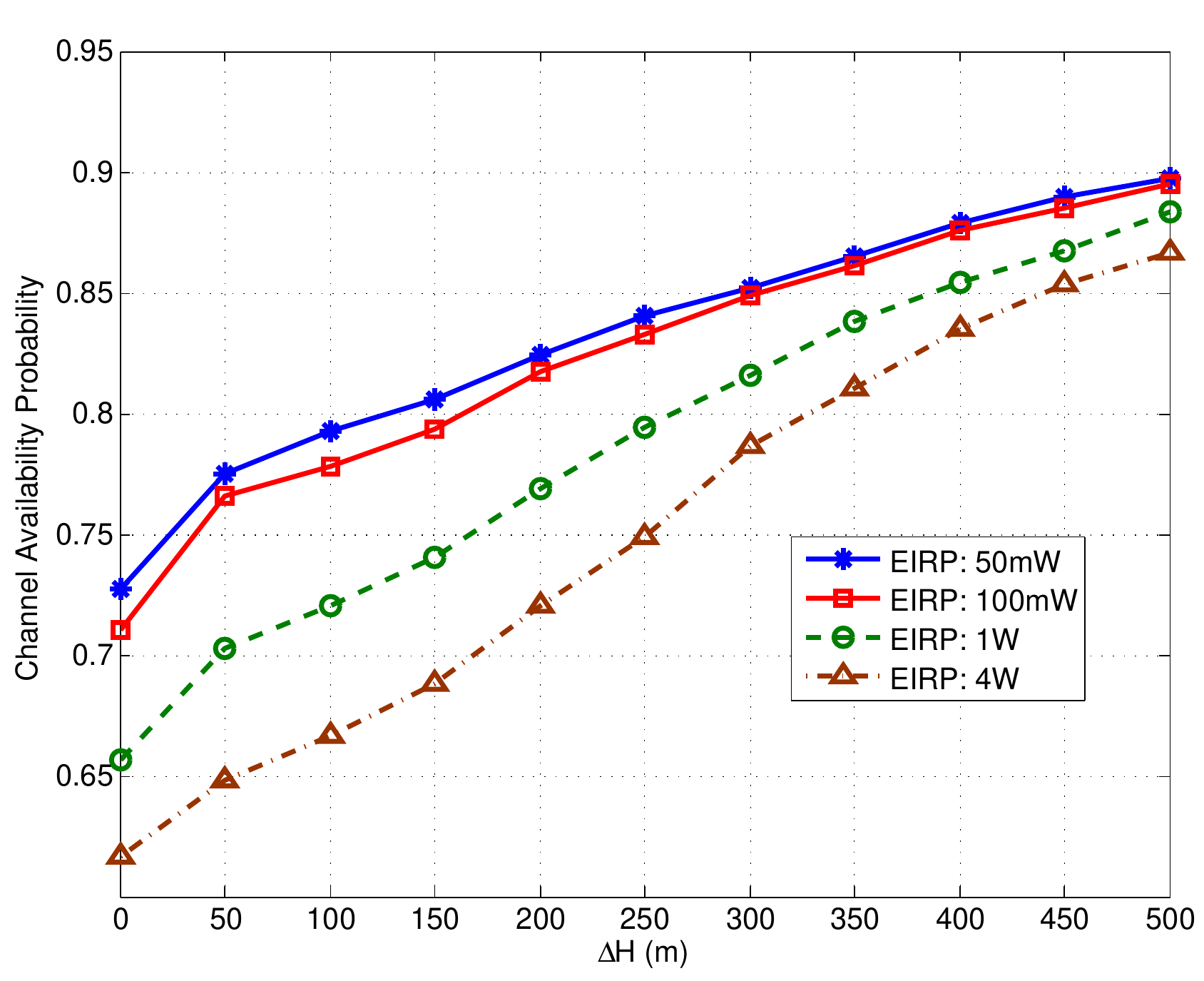}%
	\caption{Channel 14 availability probability vs. terrain irregularity parameter $\Delta H$. Results are averaged over the entire United States.}
	\label{fig:ChanPronVsDeltaH}%
\end{figure}

Fig. \ref{fig:ChanPronVsDeltaH} plots $p(\Gamma)$ versus $\Delta H$ for channel 14 and for different values of EIRP. Increasing $\Delta H$ translates to further irregularity of the area, larger values of path loss and eventually smaller protection regions. This will leave wider areas for unlicensed operation by TVBDs and therefore $p(\Gamma)$ increases. It also shows how $p(\Gamma)$ depends on the value of path loss and can vary from 60\% to 90\%.

\subsection{Cell Size}
The throughput of each user in the secondary cellular network depends on the available capacity in that cell as well as the number of users $U_R$ resuesting service in the cell. By assuming a MAC layer with efficiency of $\eta_{MAC}$, capacity per user can be formulated as 
\begin{align}
C_{User} = \frac{\eta_{MAC} \overline{C_{cell}}}{U_R}
\end{align}
The number of service requests in a cell depends on population density and cell size, $U_R\propto \rho_{Pop}A_{cell}$. Thus $U_R=\alpha\rho_{Pop}\pi r_{cell}^2$ where $\alpha$ is propotionality constant. Therefore, capacity per user can be rewritten as 
\begin{align}
C_{User} = \frac{\eta_{MAC}}{\alpha \rho_{Pop}\pi} \frac{\overline{C_{cell}}}{r_{cell}^2}
\end{align}
where the first factor is a constant and does not depend on network planning parameters. The second factor however is the normalied total capacity per area (bit/sec/$m^2$) and must be optimized to achieve the best throughput per user. The capacity per area (CPA) depends on the various parameters that we studied before as well as cell size. In the following, we will explore the behaviour of CPA versus $r_{cell}$ for which we employ (\ref{eq:AvgChanwithSINR2}) and evaluate it for various values of $r_{cell}$.

\begin{figure}[!t]%
	\centering
	\includegraphics[width=\figwidth]{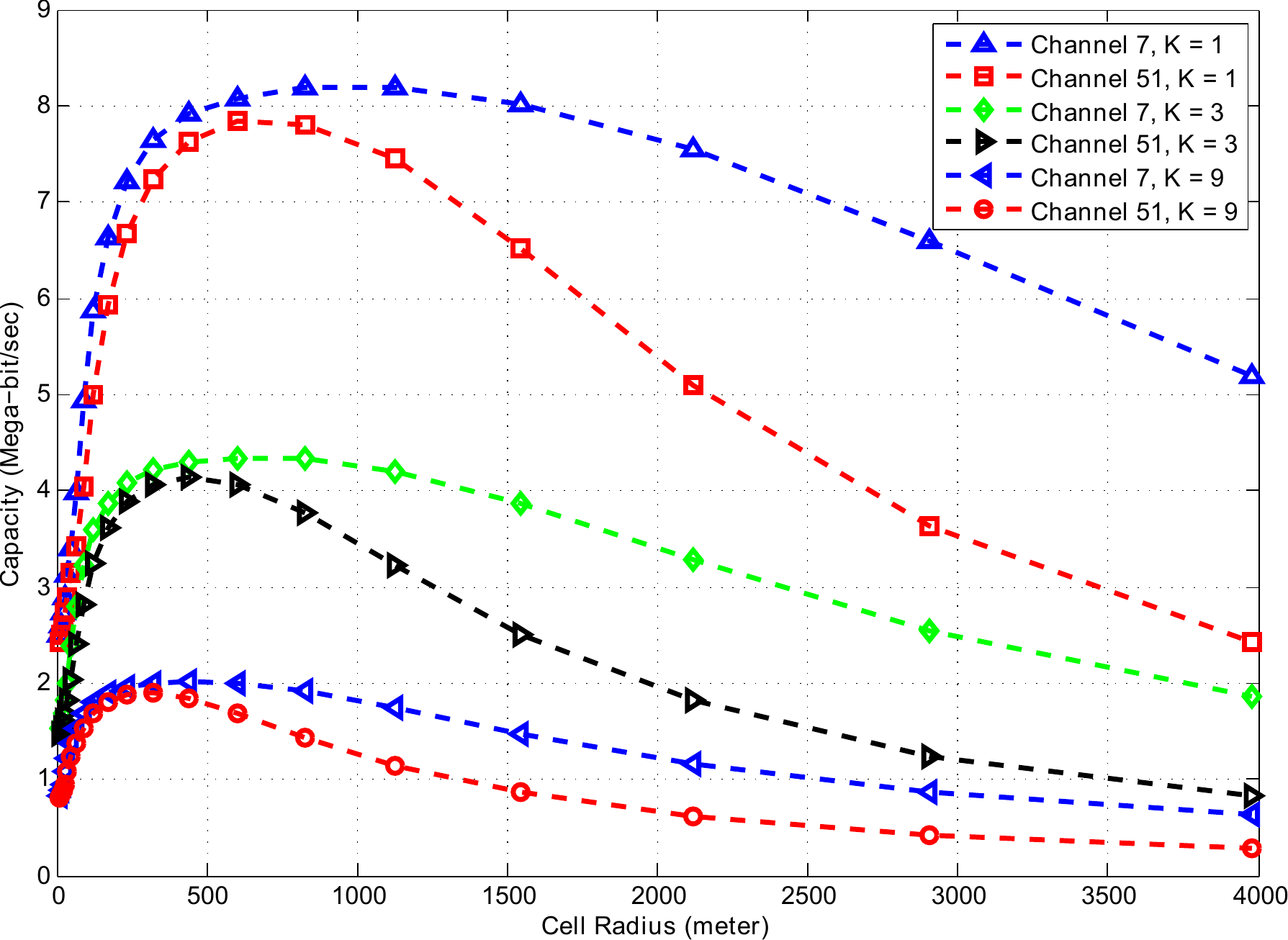}%
	\caption{Cell capacity versus cell radius for static (non-mobile) user, for various values of $K$ and channel number}
	\label{fig:StaticCapacityVsRcell}
\end{figure}

\begin{figure}[!t]%
	\centering
	\includegraphics[width=\figwidth]{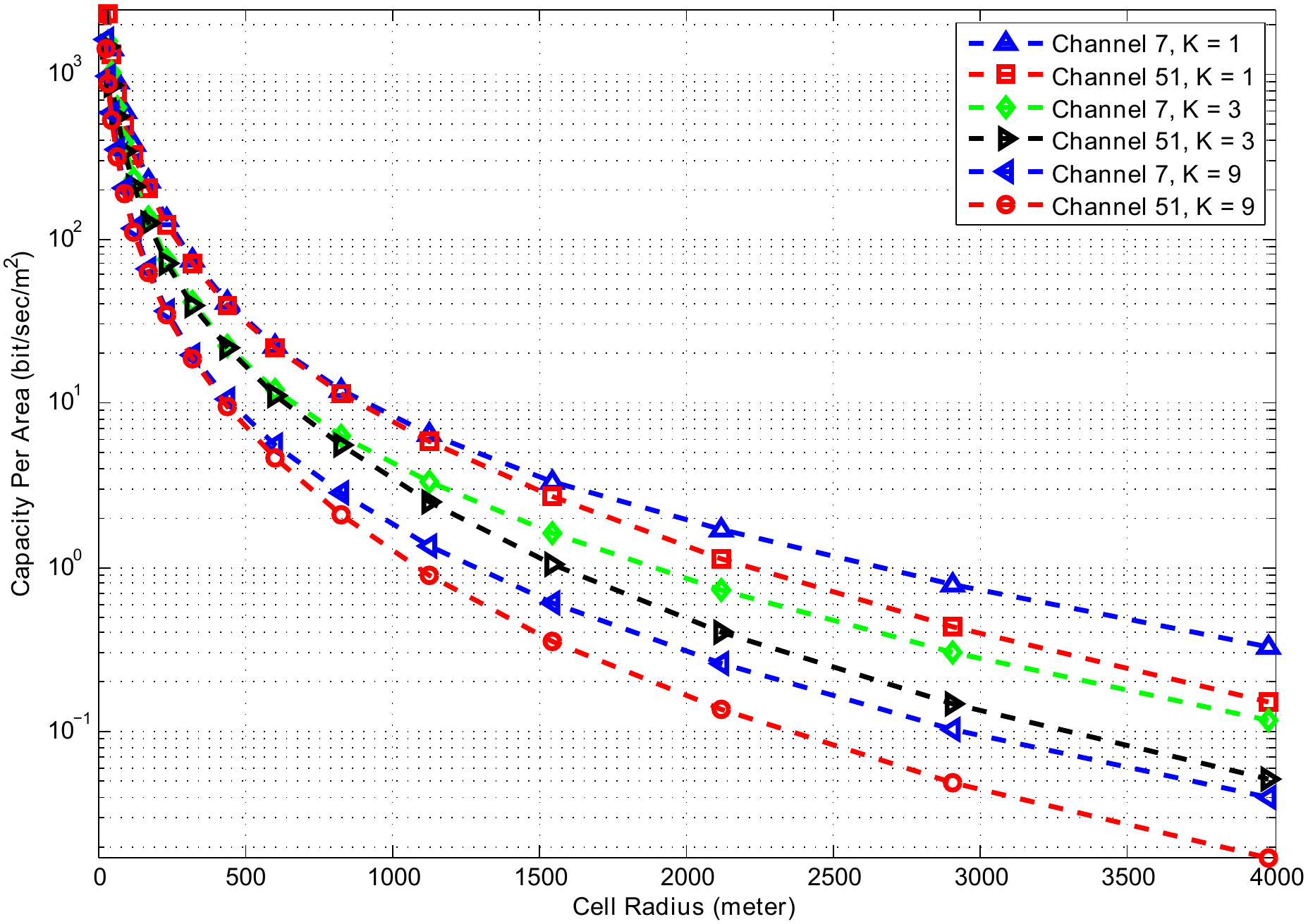}%
	\caption{Average arial capacity versus cell radius for static (non-mobile) user, for various values of $K$ and channel number}
	\label{fig:StationaryCapacityPerUserVsRcell}
\end{figure}
Fig. \ref{fig:StaticCapacityVsRcell} shows capacity of a cell on different channels (channel 7 and 51) as a function of cell radius. The trade-off between the cell size and total capacity on each channel is clear for different frequency reuse patterns, $K=1, 3, 9$ which is due to counter effect of higher received power and higher inter-cell interference. These results are averaged over different distances within the cell. The capacity per user, however, shows a different behaviour versus cell size as shows in Fig. \ref{fig:StationaryCapacityPerUserVsRcell}. The results predicts a uniform increase in capacity per user as cell radius decreases. The reason is because the number of users in the cell decreases proportional to $\frac{1}{r_{cell}^2}$. This result is interesting because we can improve the capacity of each user by deploying more base stations with reduced coverage range. This results however is valid for stationary users that are not handed off from one cell to another. 

\begin{figure}[!t]%
	\centering
	\includegraphics[width=\figwidth]{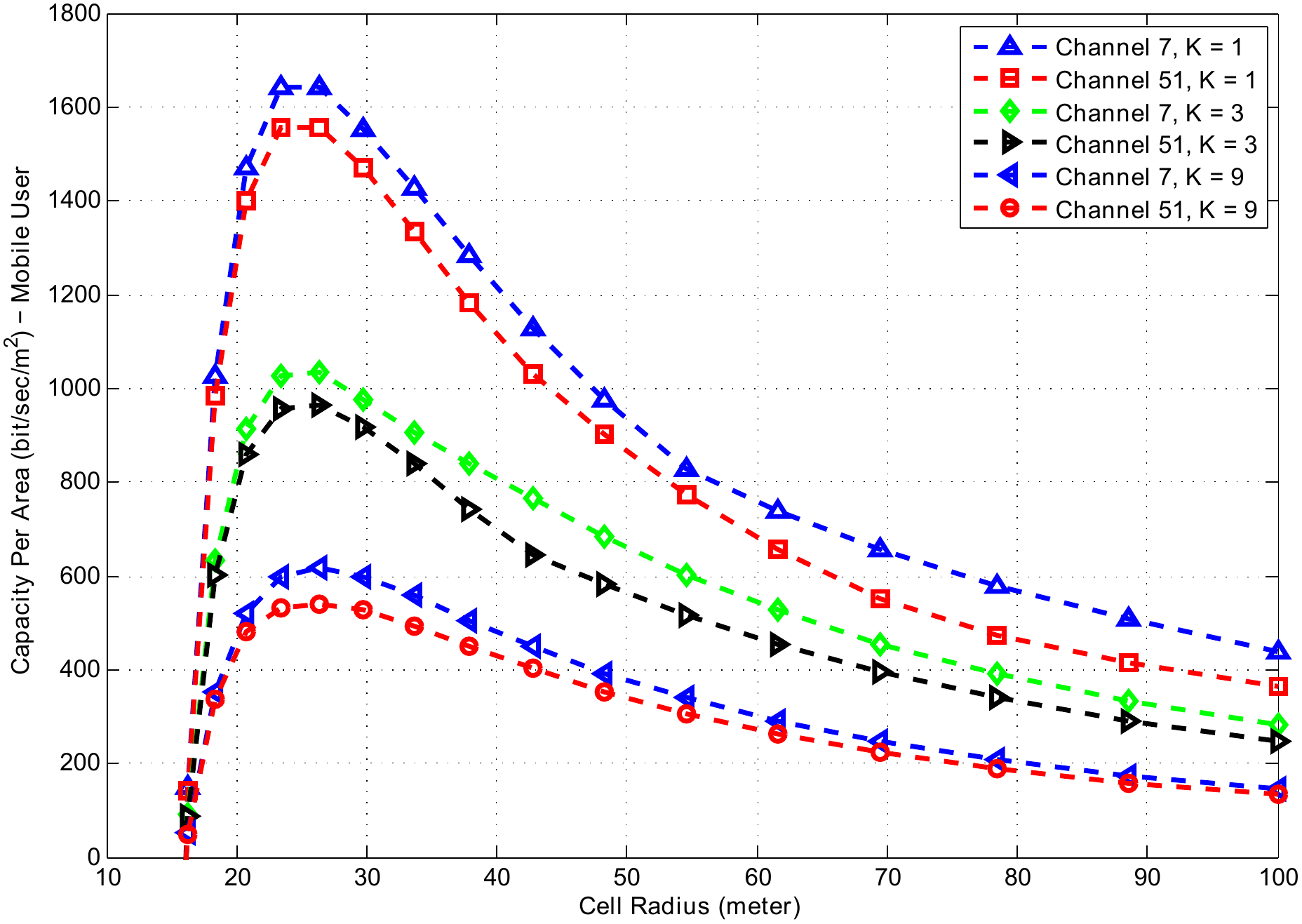}
	\caption{Average arial capacity versus cell radius for mobile user with speed of 50 Km/h, for various values of $K$}
	\label{fig:MobileCapacityVsRcell}
\end{figure}

For a mobile user that is moving from one cell to another cell, if the cell size is very small then the user has to be handed off very often. The hand-over process usually takes a certain delay overhead $\tau_{HO}$, during which user data is not transmitted to base stations. Let's assume user is moving in a straight line with speed $V$ m/s. During a duration of $t$, the number of times that the user must be handed off is proportional to $\frac{Vt}{r_{cell}}$. Therefore, the effective capacity that is experience by the mobile user is
\begin{align}
C_{Mobile} = (1-\frac{\alpha \tau_{HO} V}{r_{cell}}) C_{User}
\end{align}
with $\alpha$ being a proportionality constant. Fig. \ref{fig:MobileCapacityVsRcell} shows the result for a mobile user of speed $V$=50 km/h and hand-over delay $\tau_{HO}=1 s$. As this figure suggests, the cell size cannot be less than 25 meters because the effective capacity drops fast.

\section{Conclusion}
We introduced a detailed model for secondary network capacity that includes the key parameters in both primary and secondary networks. The model includes information from current FCC database for primary TV broadcasters and follows FCC rules for incumbent protection by considering protected contours and minimum separation distances. The proposed secondary network was arranged in a cellular layout inside the allowed regions for sharing WS channels between TVBDs.

The model described how secondary network capacity depends on TVBD system parameters such as {\em power, antenna height, device type} and also on FCC protection regulations. As noted before, current FCC rules are over protective and leave an average of $10$ channels unused\footnote{This should be modified to $6$ channels if considering the average 4 channels that are reserved for wireless microphones}, mainly due to poor design of TV receivers which cannot coexist with adjacent channel transmitters. Furthermore, we explored the effect of primary network design on secondary capacity by considering the transmission mask for TV broadcasters that extends beyond the 6 MHz channel up to two adjacent channels.

The model was used for evaluation of various trade-offs that exists between capacity versus power, antenna height, and path loss parameters to illustrate possible optimizations in future TVWS network designs. In particular, the spatial variation of TVWS network capacity - as presented in terms of average number of available channels and CDF of total number of channels - as a function of population density, deserves particular attention by beyond 4G network planners.

\appendices
\section{Protection Radius Calculation}{\label{app:pr}}
\subsection{Protected Contour $r_{PC}$}
For a given TV transmitter, protection contour $r_{PC}$ is found by finding the maximum distance where signal strength (in dBu) or equivalently signal power (in dBm) drops to minimum threshold $\Delta$, defined in Table \ref{tb:interferenceThr}. The received power $P_r$ is defined in terms of transmitted power $P_t$ as:
\begin{align}
P_r = P_t + G_t - L_{TV}(r_{PC}) + G_r = \Delta
\end{align}
where $G_t$ ($G_r$) is transmitter (receiver) antenna gain and $L_{TV}(.)$ is the path loss model for TV signals as a function of distance to transmitter. The range of protection contour highly depends on this path loss model:
\begin{align}
r_{PC} = L_{TV}^{-1}(P_t+G_t+G_r-\Delta)
\label{eq:rpc}
\end{align}
Note that protection contour is only a feature of TV transmitter and does not depend on secondary user parameters.

\subsection{Protection Region $r_p$}
Protection region is defined for every application of secondary users. For example for fixed TVBD, power limit is higher and it forces secondary users to be further away from TV transmitter than portable devices with much lower power limits. Let's assume $\gamma$ is the desired interference ratio $\gamma=\frac{\mbox{Desired Signal Power}}{\mbox{Undesired Signal Power}}$, $P_{sec}$ is the power of secondary user, and $G_{sec}$ is the antenna gain of secondary transmitter. The minimum separation distance $d_{MS}$ as shown in Fig. \ref{fig:coverage} should be such that resulting $\gamma$ is at least equal to threshold $\gamma_0$ given by Table \ref{tb:interferenceThr} for various TV channels and services.
\begin{align}
P_{sec} + G_{sec} - L_{sec}(d_{MS}) + G_r \leq \Delta - \gamma_0
\end{align}
where $L_{sec}(.)$ is the path loss model for secondary transmitter. Using this, $r_p$ is found to be:
\begin{align}
r_p = d_{MS} + L_{sec}^{-1}( P_{sec} + G_{sec} + G_r - \Delta + \gamma_0 )
\label{eq:rn}
\end{align}
This equation shows how $P_{sec}$ plays an important role in calculation of channel availability.

\begin{table}[tbp]
\caption{Desired to undesired signal ratio defined by FCC for maximum tolerable interference in various TV applications}
\label{tb:interferenceThr}
\renewcommand{\arraystretch}{1.3}
\begin{tabular}{ p{3.5cm} c c }
\toprule
\multirow{2}{*}{Type of Station} & \multicolumn{2}{ c }{Protection ratios} \\
\cline{2-3}
                                 & Channel Separation & D/U ratio (dB) \\
\hline
\multirow{3}{3.5cm}{Analog TV, Class A, LPTV, translator and booster}& Co-channel & 34 \\ \cline{2-3}
& Upper adjacent & -17 \\ \cline{2-3}
& Lower adjacent & -14 \\ \cline{2-3}
\hline\hline
\multirow{3}{3.5cm}{Digital TV and Class A}& Co-channel & 23 \\ \cline{2-3}
& Upper adjacent & -26 \\ \cline{2-3}
& Lower adjacent & -28 \\
\bottomrule
\end{tabular}
\end{table}

\begin{table}[tbp]
\caption{TV Station Protected Contours; Note that threshold values are in dBu and represents signal strength (not power)}
\label{tb:table1}
\renewcommand{\arraystretch}{1.3}
\begin{tabular}{ p{2.5cm} p{2cm} p{1.0cm} p{1.5cm}}
\toprule
\multirow{2}{*}{Type of Station} & \multicolumn{3}{ c }{Protection contour} \\
\cline{2-4}
& Channel & Contour (dBu) & Propagation curve \\
\hline
\multirow{3}{2cm}{Analog: Class A, LPTV, translator and booster}   & Low VHF (2-6)  & 47 & F(50,50) \\ \cline{2-4}
& High VHF (7-13)& 56 & F(50,50) \\ \cline{2-4}
& UHF (14-51)    & 64 & F(50,50) \\ \hline
\multirow{3}{2.5cm}{Digital: Full service TV, Class A TV, LPTV, translator and booster}   & Low VHF (2-6)  & 28 & F(50,90) \\ \cline{2-4}
& High VHF (7-13)& 36 & F(50,90) \\ \cline{2-4}
& UHF (14-51)    & 41 & F(50,90) \\ \hline
\bottomrule
\end{tabular}
\end{table}

\begin{table}[tbp]
\caption{Modified Field Strengths Defining the Area Subject to Calculation for Analog Stations}
\label{tb:AnalogThreshold}
\renewcommand{\arraystretch}{1.3}
\centering
\begin{tabular}{||c|p{7cm}||}
\hline
\hline
Channels & Defining Field Strengths, dBu, to be predicted using F(50, 50) \\
\hline
2 - 6    &            47 \\
\hline
7 - 13   &            56 \\
\hline
14 - 69  &            $64-20\log_{10}\left(\frac{615}{\mbox{channel mid frequency in MHz}}\right)$\\
\hline
\end{tabular}
\end{table}

\section{ITM Path-Loss Model}
The ITM models path loss in area mode in terms of a reference attenuation $A_{ref}$. This is the {\em median} attenuation relative to a free space signal that should be observed on the set of all similar paths during times when the atmospheric conditions correspond to a standard, well-mixed, atmosphere.

The reference attenuation is determined as a function of the distance d from the piecewise formula:
\begin{align}
A_{ref} = \left\{ \begin{array}{lcc}
									\multicolumn{2}{l}{\max[0, A_{el}+K_1d+K_2\ln(d/d_{Ls})],}  & d\leq d_{Ls}\\
									A_{ed}+m_dd,                            & \multicolumn{2}{r}{d_{Ls} \leq d\leq d_x}\\
									A_{es}+m_sd,                            & \multicolumn{2}{r}{d_x \leq d}
									\end{array}
					\right.
\label{eq:Aref}
\end{align}
where the coefficients $A_{el}$, $K_1$, $K_2$, $A_{ed}$, $m_d$, $A_{es}$, $m_s$, and the distance $d_x$ are calculated using the ITM algorithms. The three intervals defined here are called the line-of-sight, diffraction, and scatter regions. The dependency of path loss on frequency is not apparent in (\ref{eq:Aref}) but all the parameters are function of frequency\footnote{This dependency is not straightforward. They also depend on terrain irregularity parameter, Tx/Rx height, radio climate, polarization, etc.}. The total path loss is the sum of $A_{ref}$ and free space path loss which also depends of frequency:
\begin{align}
L(d) = A_{ref} + 20\log_{10}\left(\frac{4\pi df}{C}\right)
\label{eq:totalPathLoss}
\end{align}
where $C$ is the speed of light in vacuum. The irregularity parameter for an average terrain is $\Delta H = 90$; Using this, the reference attenuation for lowest/highest available frequency in TV band (channel 2 = 57 MHz, channel 51 = 695 MHz) are found as:
Channel 2:
\begin{align}
A_{ref} = \left\{ \begin{array}{lcr}
									\multicolumn{2}{l}{\max[0, -7.1+d\times 4.03e-4],}  & d\leq 94.1\text{ km}\\
									5.87+d\times 2.65e-4,          & \multicolumn{2}{r}{94.1 \text{km} \leq d\leq 159\text{ km}} \\
									38.58+d\times 5.95e-5,         & \multicolumn{2}{r}{159\text{ km} \leq d}
									\end{array}
					\right. \nonumber
\end{align}
Channel 51:
\begin{align}
A_{ref} = \left\{ \begin{array}{lcr}
									\multicolumn{2}{l}{\max[0, -17.94+d\times 5.08e-4],}  & d\leq 94\text{ km}\\
									-17.2+d\times 5.0e-4,          & \multicolumn{2}{r}{94 \text{km} \leq d\leq 136\text{ km}} \\
									42+d\times 6.56e-5,            & \multicolumn{2}{r}{136\text{ km} \leq d}
									\end{array}
					\right.
\label{eq:ArefSample2}
\end{align}

According to analysis in sec. \ref{sec:CellularNetworkStructure} to sec. \ref{sec:InterferenceModel}, determination of free channels as well as capacity calculations in TVWS highly depends on path loss model behavior. The more precise the model the less spectrum resources are wasted and the better protection is provided to TV receivers. In order to achieve some intuition about how coefficients in ITM model, (\ref{eq:Aref})-(\ref{eq:totalPathLoss}), depends on fundamental parameters such as {\em frequency}, {\em antenna height} and {\em irregularity parameter}, a simplified version of ITM model that is more specific to TVWS system parameters is provided here. The original detailed ITM model can be found in \cite{ITM:Model}.
\begin{align}
A_{ref} = \left\{ \begin{array}{lcc}
									\multicolumn{2}{l}{\max[0, A_{el}+K_1d+K_2\ln(d/d_{Ls})],}  & d\leq d_{Ls}\\
									A_{ed}+m_dd,                            & \multicolumn{2}{r}{d_{Ls} \leq d\leq d_x}\\
									A_{es}+m_sd,                            & \multicolumn{2}{r}{d_x \leq d}
									\end{array}
					\right.
\label{eq:Aref2}
\end{align}
where line-of-sight distance $d_{Ls}$ is defined as:
\begin{align}
d_{Ls} &= d_{Ls1} + d_{Ls2} = \sqrt{\frac{2h_{g,TX}}{\gamma_e}} + \sqrt{\frac{2h_{g,RX}}{\gamma_e}}
\end{align}
where $h_{g,TX}/h_{g,RX}$ are transmitter/receiver structural height and $\gamma_e$ is the Earth's curvature constant. The diffraction range parameters are defined as:
\begin{align}
m_d &= \frac{A_{\text{diff}}(d_L+4.3161X_{ae})-A_{\text{diff}}(d_L+1.3787X_{ae})}{2.7574X_{ae}} \\
A_{ed} &= A_{\text{diff}}(d_L+1.3787X_{ae}) - m_d * (d_L+1.3787X_{ae})
\end{align}
with
\begin{align}
d_L &= \sqrt{\frac{2h_{g,TX}}{\gamma_e}}e^{-0.07\sqrt{\Delta h/h_{g,TX}}} + \sqrt{\frac{2h_{g,RX}}{\gamma_e}} e^{-0.07\sqrt{\Delta h/h_{g,RX}}}\\
X_{ae} &= \left( \frac{2\pi}{\lambda} \gamma_e^2\right)^{-1/3}
\end{align}
illustrates dependency on antenna heights and terrain irregularity $\Delta H$. The diffraction function $A_{\text{diff}}(s)$ is a complex function in terms of Fresnel integral \cite{ITM:Model}. The line-of-sight coefficients in (\ref{eq:Aref2}) are simplified to $K_2=0$,
\begin{align}
K_1 = \left\{ \begin{array}{lr}
								\frac{A_{ed}+m_dd_{Ls}-A_{\text{los}}(d_0)}{d_{Ls}-d_0} & ,A_{ed}>0 \\
								\frac{A_{ed}+m_dd_{Ls}-A_{\text{los}}(d_1)}{d_{Ls}-d_1} & ,A_{ed}<0
							\end{array}
			\right.
\end{align}
where
\begin{align}
d_0 &= \min(\frac{d_L}{2}, 1.908kh_{g,RX}h_{g,TX})\\
d_1 &= \max(A_{ed}/m_d, d_L/4)
\end{align}
$A_{\text{los}}$ is also defined in \cite{ITM:Model} in terms of the `extended diffraction attenuation', $A_d$ and the `two-ray attenuation', $A_t$:
\begin{equation}
A_{\text{los}}= (1-\omega)A_d + \omega A_t
\end{equation}


\ifCLASSOPTIONcaptionsoff
  \newpage
\fi

\bibliographystyle{IEEEtran}
\bibliography{IEEEabrv,mylit}

\end{document}